\documentclass[letterpaper,prc,twocolumn,showpacs,floatfix,nofootinbib,preprintnumbers,superscriptaddress,amsmath,amssymb]{revtex4-1}

\usepackage{amsmath}           
\usepackage{amsfonts}
\usepackage{graphicx}          
\usepackage{dcolumn}           
\usepackage{bm}
\usepackage{multirow}

\usepackage{color}

\newcommand{\gras}[1]{\boldsymbol{#1}}

\begin{document}

\preprint{\fbox{\sc version of \today}}

\title{Description of Induced Nuclear Fission with Skyrme Energy
Functionals: I. Static Potential Energy Surfaces and Fission Fragment Properties}

\author{N. Schunck}
\affiliation{Physics Division, Lawrence Livermore National Laboratory, Livermore, CA 94551, USA}

\author{D. Duke}
\affiliation{School of Computing, University of Leeds, UK}

\author{H. Carr}
\affiliation{School of Computing, University of Leeds, UK}

\author{A. Knoll}
\affiliation{Argonne National Laboratory, USA}

\date{\today}

\begin{abstract}
Eighty years after its experimental discovery, a description of induced nuclear
fission based solely on the interactions between neutrons and protons and
quantum many-body methods still poses formidable challenges. The goal of this
paper is to contribute to the development of a predictive microscopic framework
for the accurate calculation of static properties of fission fragments for hot
fission and thermal or slow neutrons. To this end, we focus on the
$^{239}$Pu(n,f) reaction and employ nuclear density functional theory with
Skyrme energy densities. Potential energy surfaces are computed at the
Hartree-Fock-Bogoliubov approximation with up to five collective variables. We
find that the triaxial degree of freedom plays an important role, both near the
fission barrier and at scission. The impact of the parameterization of the
Skyrme energy density and the role of pairing correlations on deformation
properties from the ground-state up to scission are also quantified. We
introduce a general template for the quantitative description of fission
fragment properties. It is based on the careful analysis of scission
configurations, using both advanced topological methods and recently proposed
quantum many-body techniques. We conclude that an accurate prediction of
fission fragment properties at low incident neutron energies, although
technologically demanding, should be within the reach of current nuclear
density functional theory.
\end{abstract}


\maketitle

%
%
%

\section{Introduction}
\label{sec-introduction}

The accurate description of neutron-induced fission is particularly important
to address present challenges in the areas of energy production, nuclear
waste disposal or national security applications. Many of these applications
require a detailed knowledge of fission fragment properties such as their
charge, mass, and relative yields, their total kinetic energy, their total
excitation energy, etc. The fission spectrum, i.e. the number and
characteristics of both pre- and post-scission neutrons and gammas, often needs
to be known within a few percent accuracy. In many fissile or fissionable
nuclei of interest, experimental measurements are not possible and theoretical
simulations of the fission process are therefore necessary.

The central idea in the theoretical description of induced fission remains that
of Bohr and Wheeler \cite{(Boh39)}: fission is modeled as a two-step process
where the incident neutron first fuses with the target to form a compound
nucleus (in an excited state), which then breaks into two or more fragments.
These fragments will themselves decay to their respective ground-state. Based
on this hypothesis, powerful toolkits have been developed over the years to
reproduce fission data: Monte-Carlo schemes are used to simulate the
deexcitation of fission fragments after scission
\cite{(Vog13),(Bec13),(Vog12),(Vog11),(Tal11),(Vog09),(Ran09)};
reaction models focus on explaining the characteristic features of the fission
spectrum such as fission isomers, collective structures, resonances, etc.
\cite{(Bjo80)}; nuclear structure models provide basic information on the
fission fragments and the fission process itself, such as fission barrier
heights, charge, mass, and energy distributions. Many results have been
obtained using the macroscopic-microscopic approach to nuclear structure
\cite{(Mol95),(Bra72)} and its dynamical extensions using either the general
Langevin equations \cite{(Nad12),(Nad07)} or their restriction to Brownian
motion \cite{(Ran11a),(Ran11b)}. This approach is complemented by various
scission point models, the goal of which is to simulate the actual break-up of
the nucleus at large elongations \cite{(Wil76)}.

This semi-phenomenological framework has been very successful in explaining and
reproducing numerous features of the fission process; see, e.g.
Refs.~\cite{(Ich13),(Mol12),(Ran11a),(Ran11b),(Mol09),(Ich09)} for recent
applications. Nevertheless, a truly predictive theory of fission should
ultimately be based on a detailed account of the nuclear forces between protons
and neutrons combined with the use of standard many-body methods of quantum
physics. In principle, several approaches can meet these requirements. For
example, functional integral methods are fully quantum mechanical approaches
that include quantum dissipation and fluctuations \cite{(Neg88),(Ker81)}. Their
implementation, however, requires computing resources that far exceed those
available to the current generation of supercomputers. On paper, nuclear
density functional theory (DFT) represents an excellent compromise between
microscopic content and actual feasibility. In particular, DFT lends itself 
particularly well to separating nuclear excitations into fast intrinsic and 
slow collective excitations \cite{(Les13),(Ben03)}. This distinction is 
especially useful in the context of low-energy nuclear fission, which has 
timescales of the order of $10^{-19} - 10^{-20}$ seconds, i.e. two to three 
orders of magnitude slower than typical single-particle excitations. Such a 
separation is the central assumption of the time-dependent generator 
coordinate method (TDGCM), which provides an effective, quantum-mechanical 
method to compute fission fragment yields \cite{(You12a),(You12b),(Gou05),(Gou04)}.

In spite of its advantages, the proper application of nuclear DFT to
the problem of nuclear fission still requires tremendous computational
resources, especially in the determination of accurate multi-dimensional
potential energy surfaces. In the past, computer limitations imposed artificial
constraints on the theory, such as the use of small model spaces, schematic
interactions, or a reduced number of collective variables. It is only recently
that the first systematic, large-scale, and accurate simulations of nuclear
fission have been made possible. Most recent efforts have focused on
spontaneous fission in actinide and superheavy nuclei, and quantities such as
barriers and lifetimes; see e.g.
Refs.~\cite{(Rod14),(McD13),(War12),(Abu12),(War11),(Abu10),(Sta09),(Pei09)}
for a selection of recent results. In contrast, there have been comparatively
fewer publications on the topic of induced fission
\cite{(You11),(You09),(Dub08),(Gou05),(Gou04)}.

This paper is the first in a series of articles focusing on the microscopic
description of induced fission within the framework of nuclear density
functional theory with Skyrme energy densities. As such, it should be
considered as an intermediate step in the long-term effort to achieve a
predictive theory of fission. The specific goals of this first paper are (i) to
provide a comprehensive mapping of deformation properties of $^{240}$Pu, (ii)
to give a detailed and quantitative analysis of the role of triaxiality in
fission calculations, (iii) to assess the dependence of calculations on the
parametrization of the functional, and (iv) to establish a template for
the calculation of fission fragment properties. In several aspects, this study is both
a continuation and an extension of the general description of induced fission
developed over the years at the Commissariat \`{a} l'Energie Atomique in France
and Lawrence Livermore National Laboratory in the USA; Cf., for example, 
Refs.~\cite{(Ber84),(Ber86),(Ber89),(Gou04),(Gou05),(Dub08),(You09),(You09a),(You11),(Dub12),(You12a),(You12b),(You13)}.

Section \ref{sec-theory} contains a brief reminder of the nuclear density
functional approach to induced fission, Skyrme functionals and the practical
implementation of DFT. Section \ref{sec-pathway} focuses on the static
potential energy surfaces in $^{240}$Pu, which is the compound nucleus formed
in $^{239}$Pu(n,f), and their dependence on the parametrization of the Skyrme
and pairing functionals. Section \ref{sec-scission} presents a detailed
analysis of the identification of the scission point based both on topological
methods and the concept of quantum localization, and provides estimates of
fission fragment properties for the most probable fission.


\section{Theoretical Framework}
\label{sec-theory}

Our theoretical approach is based on the local density approximation of the
energy density functional (EDF) theory of nuclear structure. The next few
sections review the basic ingredients of the EDF theory pertaining to the
description of nuclear fission.


\subsection{Density Functional Theory Approach to Induced Fission}
\label{subsec-fission}

In the context of nuclear fission, the ultimate goal of nuclear density
functional theory is to provide a comprehensive and accurate description of
both the fissioning nucleus (half-lives, fission probability) and the fission
fragments (mass and charge distributions, excitation energy, yields, etc.)
based on the best knowledge of nuclear forces and quantum many-body techniques.

Density functional theory of nuclei is a mature field with numerous
applications in low-energy nuclear physics and nuclear astrophysics
\cite{(Ben03),(Sto07)}. The central assumption of the approach is that atomic nuclei
can be described accurately by an effective energy density $\mathcal{H}$, which
is a functional of the one-body density matrix and the pairing tensor -- since
pairing correlations play an essential role in low-energy nuclear structure.
This energy density may or may not be derived from an effective pseudopotential
$\hat{V}_{\text{eff}}$. In practice, most applications of DFT so far have used
either the Skyrme or Gogny energy density, which are indeed derived from an
effective two-body pseudopotential, of zero range for Skyrme and finite range
for Gogny. The coupling constants of the energy density are free parameters to
be determined, usually on global observables such as atomic masses, r.m.s.
radii, nuclear matter properties, etc.; See, e.g., Refs.~\cite{(Kor10),(Kor12),(Kor14)} 
for recent applications.

For the specific case of induced fission, two additional hypotheses underpin
the DFT approach:
\begin{enumerate}
\item One can identify a set of collective degrees of freedom $\gras{q}$ that
drive the dynamics of the fission process. The most important of these
collective degrees of freedom are related to the nuclear shape, although
additional collective variables related, e.g., to the pairing channel, could 
be introduced \cite{(Rod14)}. The collective degrees of freedom might be 
considered as free parameters of the theory, although the variational nature 
of DFT ensures that the more collective variables there are, the better the 
accuracy is. 
\item The transition between the compound nucleus and fully independent fission
fragments can be controlled by the introduction of scission configurations.
Without this additional constraint, the short range of nuclear forces combined
with the variational nature of DFT would always yield fission fragments in
their ground-state configurations, so that the total energy of the system be
minimal. This is contrary to experimental data, which shows that fission
fragments can be excited.
\end{enumerate}
We note that these two assumptions are reminiscent of semi-phenomenological
approaches to fission, in particular scission point models using inputs from
macroscopic-microscopic potential energy surfaces \cite{(Wil76)}. The main
difference is that DFT is built onto a unique energy density that
simultaneously determines bulk and shell effects, the collective inertia, and
the dynamics of the problem in a unifying quantum-mechanical framework.

Based on the aforementioned hypotheses, the full DFT description of
{\em induced} fission relies on the following multi-step approach
\begin{enumerate}
\item Static properties of the fissioning nucleus are computed as a function of
the collective degrees of freedom $\gras{q}$. These potential energy surfaces
are obtained by solving the DFT equations, which most often take the form of
the Hartree-Fock-Bogoliubov equations with constraints. This step can be viewed
as the construction of an adequate basis made of those nuclear many-body states
that are the most relevant for the fission process.
\item Scission configurations are then identified on the potential energy
surface, based on some criteria. It is precisely the purpose of this paper to
discuss in details some of these criteria.
\item Fission fragment properties are obtained by solving the many-body
time-dependent Schr\"{o}dinger equation under the general assumptions of DFT,
namely that the ground-state takes the form of a Slater determinant (HF) or,
when pairing correlations are included, of a quasi-particle vacuum (HFB). This
could be done ``directly'' with the time-dependent Hartree-Fock (TDHF) theory,
and its extension with pairing, the time-dependent Hartree-Fock-Bogoliubov
(TDHFB) theory \cite{(Sim14),(Neg78)}. Alternatively, one may  use the basis 
of many-body states generated in step 1 to formulate a collective, 
time-dependent, Schr\"{o}dinger-like equation: this is the essence of the 
time-dependent generator coordinate method (TDGCM) \cite{(Ber84),(Ber86),(Ber89)}. 
In practice, only the TDGCM has been applied to the study of fission fragment
distributions so far \cite{(Gou04),(Gou05),(You12a),(You12b),(You13)}.
\end{enumerate}
One should emphasize that, strictly speaking, the DFT description of fission
requires {\em all} of the aforementioned steps. In particular, while potential
energy surfaces can provide valuable inputs to reaction models, or even be used
to compute pseudo-experimental quantities such as fission barriers, they are,
in reality, only an auxiliary basis used to compute the fission fragment
properties in the TDGCM. In this paper, the focus is on selected topics
pertaining to the {\it static} aspects of fission. We leave the calculation of
fission fragment properties, yields, and distributions to a forthcoming paper.


\subsection{Skyrme Energy Functional}
\label{subsec-skyrme}

In the local density approximation of the EDF theory, the energy of
the nucleus is given as the integral over space of the Hamiltonian density
$\mathcal{H}(\gras{r})$, which is itself a functional of the one-body density
matrix $\rho$ and pairing tensor $\kappa$,
\begin{equation}
E = \int d^{3}\gras{r}\; \mathcal{H}(\gras{r})
\end{equation}
The Hamiltonian density is built out of a kinetic energy density term, a
potential energy density $\chi_{t}$, and a pairing energy density
$\tilde{\chi}_{t}$
\begin{equation}
\mathcal{H}[\rho, \kappa] = \frac{\hbar^{2}}{2m}\tau(\gras{r})
+ \sum_{t=0,1}\chi_{t}(\gras{r}) + \sum_{t=0,1}\tilde{\chi}_{t}(\gras{r}),
\end{equation}
where $\tau(\gras{r})$ is the kinetic energy density, and the index $t$ refers
to the isoscalar ($t=0$) or isovector ($t=1$) component of the potential energy
density, see Ref.~\cite{(Per04)} and references therein. In this work, the
potential energy density is obtained from the zero-range Skyrme pseudopotential
\cite{(Vau72)}. We employed three different parametrizations of the Skyrme EDF:
(i) The SkM* parametrization \cite{(Bar82)} remains a standard in fission
calculations with Skyrme EDFs, see, e.g.
Refs.~\cite{(McD13),(Sta13),(War12),(Nha12),(Sta09),(She09),(Pei09),(Bon06)}
for some recent applications. Since the parameters of the pseudopotential were
explicitly adjusted to fission barrier heights, it is believed to have good
deformation properties; (ii) The UNEDF0 EDF is a recent parametrization of the
Skyrme energy density that gives a very good agreement with nuclear masses
\cite{(Kor10)} but was shown to have unrealistic deformation properties
\cite{(Kor12),(Nik11)}: we use it only to study the impact of model parameters
on fission observables; (iii) The UNEDF1 EDF was obtained by extending the
optimization protocol of UNEDF0 to include selected data on fission isomers
\cite{(Kor12)}. It offers an excellent compromise between predictive power
(limited amount of data used in the fit) and overall quality.

In this work, pairing correlations are treated at the Hartree-Fock-Bogoliubov
(HFB) approximation \cite{(RS80w)}. The pairing energy density $\tilde{\chi}$
is a functional of the pairing tensor $\kappa$, or equivalently
of the pairing density $\tilde{\rho}$ \cite{(Ben03)}. It is
derived from a density-dependent contact pairing interaction with mixed
volume-surface character \cite{(Dob02)},
\begin{equation}
\hat{V}_{\text{pair}}(\gras{r},\gras{r'}) = V_{0}^{(n,p)}
\left[ 1 - \frac{1}{2}\frac{\rho(\gras{r})}{\rho_{c}} \right]
\delta(\gras{r}-\gras{r'}),
\label{eq:pairing}
\end{equation}
with $V_{0}^{(n,p)}$ the pairing strength for neutrons (n) and protons (p),
and $\rho_{c} = 0.16$ fm$^{-3}$ the saturation density. The energy cut-off
was set at $E_{\text{cut}} = 60$ MeV. For our calculations with the SkM* EDF,
we adjusted $V_{0}^{(n)}$ and $V_{0}^{(p)}$ locally on the 3-point odd-even
mass difference in $^{240}$Pu. This gave $V_{0}^{(n)} = -265.25$ MeV and
$V_{0}^{(p)} = -340.06$ MeV. In the case of UNEDF0 and UNEDF1, the value of the
pairing strengths $V_{0}^{(n,p)}$ is fixed by the parametrization; in addition,
calculations with these two functionals are performed using an approximate
formulation of the Lipkin-Nogami prescription \cite{(Sto03),(Sto07)}.

The nuclear shape is characterized by the expectation value $q_{\lambda\mu}$
of the multipole moment operators $\hat{Q}_{\lambda\mu}$ on the HFB vacuum.
We will also employ the expectation value of the so-called Gaussian neck
operator,
\begin{equation}
\hat{Q}_{N} = e^{-\left( \frac{z - z_{N}}{a_{N}}\right)^{2}},
\end{equation}
which gives an estimate of the number of particles in the region centered
around the point $z_{N}$ \cite{(Ber90),(War02),(You09)}. We chose the range
$a_{N} = 1.0$ fm. The collective space of nuclear fission is defined as the
ensemble of constraints imposed on the HFB solution. In this work, we will
consider the following constraints, either alone or in combinations: elongation
$\hat{Q}_{20}$, degree of triaxiality $\hat{Q}_{22}$, mass asymmetry
$\hat{Q}_{30}$, neck thickness $\hat{Q}_{40}$ and neck size $\hat{Q}_{N}$.
These collective variables will be denoted generically by
$\gras{q} = (q_{1},\cdots,q_{N})$. Constrained HFB solutions are obtained by
using a variant of the linear constraint method, in which Lagrange parameters
are updated based on the cranking approximation of the random phase
approximation (RPA) matrix \cite{(Ber80),(You09),(Sch11)}.


\subsection{DFT Solver and Numerical Precision}
\label{subsec-numerics}

All calculations were performed with the DFT solvers HFODD \cite{(Sch11)} and
HFBTHO \cite{(Sto12)}. Both solvers implement the HFB equations with Skyrme
functionals in the one-center harmonic oscillator (HO) basis. The program
HFBTHO assumes axial and time-reversal symmetry, while HFODD breaks all
symmetries.

In Cartesian coordinates, the three-dimensional HO basis is characterized by
its frequency $\omega_{0}^{3} = \omega_{x}\omega_{y}\omega_{z}$, the maximum
oscillator number $N_{\text{max}}$, the total number of basis states
$N_{\text{states}}$, and the deformation $\beta_{2}$, which accounts for the
different frequencies in each Cartesian direction.

The largest driver of basis truncation errors is the size of the basis
\cite{(Sch13a)}. In this work, we fixed $N_{\text{max}} = 31$ and
$N_{\text{states}} = 1100$. The large $N_{\text{max}}$ value ensures that
high-lying intruder orbitals that drive deformation are included up to the
largest deformation; the cut-off in the number of states is essentially imposed
by the physical limits on the memory available and CPU time taken by the
calculations.

At the large elongations encountered in the description of fission, the
truncation of the HO model space results in a strong dependence of the HFB
calculations on the basis frequency and deformation. Based on several
experiments, we assume the oscillator frequency $\omega_{0}$ and basis
deformation $\beta_{2}$ vary with the requested expectation value $q_{20}$ of
the axial quadrupole moment $\hat{Q}_{20}$ according to
\begin{equation}
\omega_{0} =
\left\{ \begin{array}{l}
0.1 \times q_{20}e^{-0.02 q_{20}} + 6.5 \text{MeV}\ 
\text{if}\ |q_{20}| \leq 30 \text{b} \\
8.14 \text{MeV} \ \text{if}\ |q_{20}| > 30 \text{b}
\end{array}
\right.
\label{eq:omega}
\end{equation}
and
\begin{equation}
\beta = 0.05\sqrt{q_{20}}
\label{eq:beta}
\end{equation}
This choice largely mitigates basis truncation effects up to the scission
point, where we empirically estimate the error on the total energy to be of the
order of 2-3 MeV \cite{(Sch13a)}.

From the estimates given above, it should be clear that accurately capturing
the physics of fission with one-center bases is extremely challenging. Recent
studies of convergence properties in the HO basis have pointed to the existence
of more reliable extrapolation methods \cite{(Coo12),(Fur12)}. Translating
these results in the context of DFT may not be straightforward: contrary to the
{\it ab initio} approach, the effective Hamiltonian of Skyrme EDFs depends on
the density, hence on the model space. The alternatives to the one-center HO
basis all have limitations of their own. Codes using the two-center HO basis
\cite{(Dub08),(Gou05),(Ber84)}, where basis functions must be
re-orthogonalized, or the coordinate-space representation of quasiparticle
wave-functions \cite{(Pei08)} do not currently include triaxiality. Lattice
representations generate large amounts of data \cite{(Bon05)}. A promising
alternative based on multi-resolution wavelet representation of HFB
wave-functions \cite{(Fan09)} remains in its infancy and may incur a high cost
of a single HFB calculation. As we progress in our understanding of fission
mechanisms, however, it will become more and more necessary to improve the
numerical precision of DFT solvers.


\section{Static Deformation Properties of $^{\boldsymbol{240}}$Pu}
\label{sec-pathway}

In this section, we discuss the features of the static potential energy
landscape of $^{240}$Pu. In particular, our goals are to (i) discuss and
highlight the role of several shape collective variables, (ii) assess more
specifically the effect of triaxiality on the barriers and beyond scission,
(iii) quantify the effect of the parameterization of the energy density
on predictions of static fission pathways.


\subsection{Overview of the Potential Energy Surface of $^{\boldsymbol{240}}$Pu}
\label{subsec-PES}

\begin{figure}[!ht]
\center
\includegraphics[width=\linewidth]{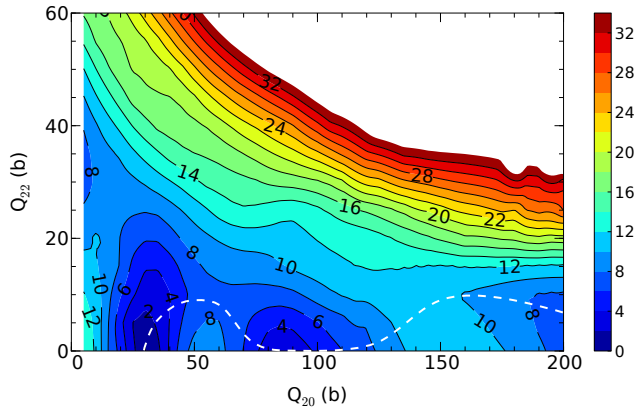}
\caption{(color online) Two-dimensional potential energy surface of $^{240}$Pu
in the $(q_{20},q_{22})$ plane for the SkM* EDF. The energy is relative to the
ground-state value. The dashed line represents the symmetric, triaxial
least-energy pathway.
}
\label{fig:PES_2D_q20q22}
\end{figure}

We begin by presenting a set of two-dimensional potential energy surfaces (PES)
that provide useful information on the local topography of the total energy in
the 4-dimensional collective space introduced at the end of
Sec.~\ref{subsec-skyrme}. In Fig.~\ref{fig:PES_2D_q20q22}, we plot the total
HFB energy as a function of the quadrupole degrees of freedom in the vicinity
of the ground-state and the fission barriers. In this calculation, the octupole
moment was set to 0 (symmetric path), and the hexadecapole moment was left
unconstrained. The ground-state,
$(q_{20}, q_{22}) \approx (35\,\text{b}, 0\,\text{b})$, and
fission isomer, $(q_{20}, q_{22}) \approx (80\,\text{b}, 0\,\text{b})$, are
clearly visible, as well as the lowering of the first fission barrier owing to
triaxiality. Although less visible in the contour map, the second barrier is
also slightly triaxial. We will quantify the effect of triaxiality on the
least-energy fission pathway in more details in Sec.~\ref{subsec-triaxial}.

\begin{figure}[!ht]
\center
\includegraphics[width=\linewidth]{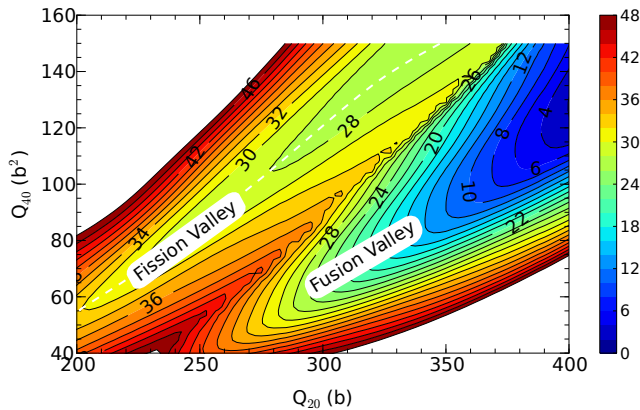}
\caption{(color online) Two-dimensional potential energy surface of $^{240}$Pu
in the $(q_{20},q_{40})$ plane for the SkM* functional. The energy is relative
to -1830 MeV. The dashed line represents the least-energy pathway.
}
\label{fig:PES_2D_q20q40}
\end{figure}

Next, we show in Fig.~\ref{fig:PES_2D_q20q40} the potential energy surface in
the $(q_{20},q_{40})$ plane. The well-known fusion (in the right-hand side of
the figure) and fission (in the left-hand side) valleys are clearly visible. We
note that the barrier between the two valleys is smaller in our Skyrme SkM*
calculations than, e.g. for the Gogny D1S functional \cite{(You09)}. For the
hot fission process, the least-energy fission pathway starts from the g.s.
and follows the fission valley until the barrier between the fission and
fusion valleys vanishes.

\begin{figure}[!ht]
\center
\includegraphics[width=\linewidth]{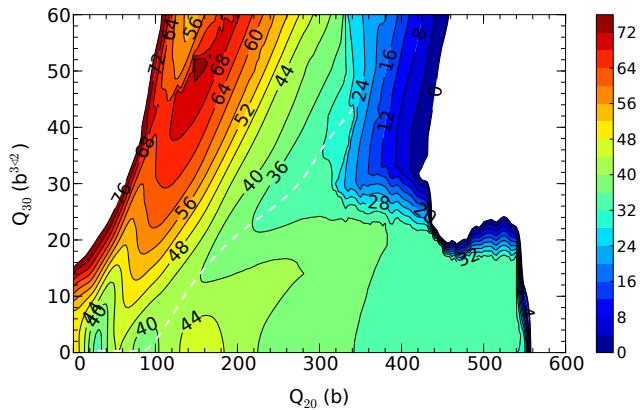}
\caption{(color online) Two-dimensional potential energy surface of $^{240}$Pu
in the $(q_{20},q_{30})$ plane for the SkM* functional. The energy is given
relative to -1840 MeV. The dashed line represents the least-energy pathway.
}
\label{fig:PES_2D_q20q30}
\end{figure}

Finally, we probe the mass asymmetry degree of freedom. In
Fig.~\ref{fig:PES_2D_q20q30}, we show the potential energy surface in the
$(q_{20},q_{30})$ plane. Since the fission fragment mass distribution of
$^{240}$Pu is known to be asymmetric, this degree of freedom is among the most
important for a quantitative description of induced fission. This calculation
is by far the largest, as we have to cover all the collective space from
symmetric fission (up to $q_{20} \approx 550\, \text{b}$ to highly asymmetric
fission (up to $q_{30}\approx 70\, \text{b}^{3/2}$). In addition, accurate
prediction of the fission fragment properties (charge and mass distributions,
kinetic energies, etc.) require the good identification of the scission region,
hence a relatively dense mesh.

The figure shows the least-energy fission pathway, which goes from about
$q_{20}\approx 100\, \text{b}$ and $q_{30} =  0\, \text{b}^{3/2}$ and exits
near $q_{20}\approx 345\, \text{b}$ and $q_{30} = 40\, \text{b}^{3/2}$. We note
that there is another fission valley that starts directly from the ground-state
and exits at small elongation but a very large asymmetry of about
$q_{30} > 60\, \text{b}^{3/2}$. This exotic, very asymmetric, fission channel
corresponds to cluster radioactivity and was discussed recently in
Ref.~\cite{(War11)}.

We also emphasize that the PES of Fig.~\ref{fig:PES_2D_q20q30} exhibits clear
signs of discontinuities, especially (but not exclusively) in the region
$ 300 < q_{20} < 550 \, \text{b}$ and $q_{30} \approx 20\, \text{b}^{3/2}$. As
discussed in detail in Ref.~\cite{(Dub12)}, these discontinuities are the
consequence of using the self-consistent procedure in a truncated collective
space of finite size: only a limited number of collective variables are
explicitly constrained, which produces these numerical artifacts. Such
discontinuities, however, provide also great physical insight since they
``automatically'' signal where collective degrees of freedom are missing for
the proper description of the process.


\subsection{Fission Pathway of Least-Energy}
\label{subsec-triaxial}

From this section on, we will focus exclusively on the least-energy fission
pathway. It is defined as the pathway connecting the ground-state to the point
of scission, along which the energy remains a local minimum in the full
collective space. It was shown recently that the dynamic fission pathway, as
obtained from the minimization of the collective action together with the
proper treatment of the collective inertia, is very close to the least-energy
pathway \cite{(Sad13)}. The latter is, therefore, a good approximation of the
most probable fission path.

\begin{figure}[!ht]
\center
\includegraphics[width=\linewidth]{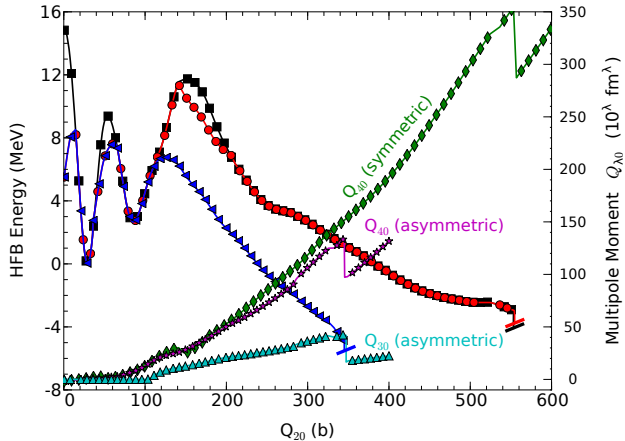}
\caption{(color online) Energy along the least-energy fission pathway in
$^{240}$Pu for the SkM* EDF: axial symmetric path (black squares), triaxial
symmetric path (red circles), triaxial asymmetric path (blue triangles). Energy
curves are given relative to the ground-state. The value of the octupole and
hexadecapole moments are also shown along the symmetric and asymmetric paths.
}
\label{fig:PES_1D_energy}
\end{figure}

In Fig.~\ref{fig:PES_1D_energy}, we superimpose the energy along the
least-energy energy fission pathway in three different scenarios: (i) symmetric
($q_{30}=0\, \text{b}^{3/2}$) fission with no triaxiality
($q_{22}=0\, \text{b}$, or $\gamma = 0^{\mathrm{o}}$), (ii) symmetric fission
with triaxiality, (iii) asymmetric fission with triaxiality. In scenario (ii),
we introduced a constraint on the expectation value of $\hat{Q}_{22}$ during
the first few iterations of the self-consistent procedure, before completely
releasing this constraint: this enabled the nucleus to jump into a triaxial
region in the case there would have been a small barrier between the axial and
triaxial solutions; finally, in scenario (iii), the same methodology was
repeated for the octupole degree of freedom $\hat{Q}_{30}$.

It is well-known that including triaxiality lowers the first barrier
\cite{(Lar72),(Gir83),(Sta05)}. It also lowers the second barrier, but only
along the symmetric fission path. We find that the degree of triaxiality is
large at the first barrier, $\gamma \approx 32^{\text{o}}$ and remains
significant in the second barrier, $\gamma \approx 15^{\text{o}}$. As seen
from Fig.~\ref{fig:PES_1D_energy}, the first barrier is lowered by
approximately 2 MeV when triaxiality is included. We note that both the
octupole and hexadecapole moment vary relatively smoothly along the path.

A clear deficiency of the SkM* functional is that the first fission barrier
height is $E_{A} \approx 7.64$ MeV, which is about 1.6 MeV higher than the
empirical barrier \cite{(Cap09),(Smi93)}. However, predictions of SkM* are in
the same ballpark as those of competing models \cite{(McD13)}. In addition, the
experimental uncertainty for the fission barrier (which is not an observable)
is usually estimated to be of the order of 1 MeV. One should, therefore, be
satisfied with an overall reproduction of barriers within 1 - 2 MeV of the
empirical value. Similarly, the fact that the one-neutron separation energy of
$^{240}$Pu computed with SkM* is $S_{n}= 7.04$ MeV, which is lower than the top
of the barrier and (unrealistically) implies that $^{239}$Pu is not fissile,
should not be cause of special concern because of the uncertainties on the
fission barriers.

\begin{figure}[!ht]
\center
\includegraphics[width=\linewidth]{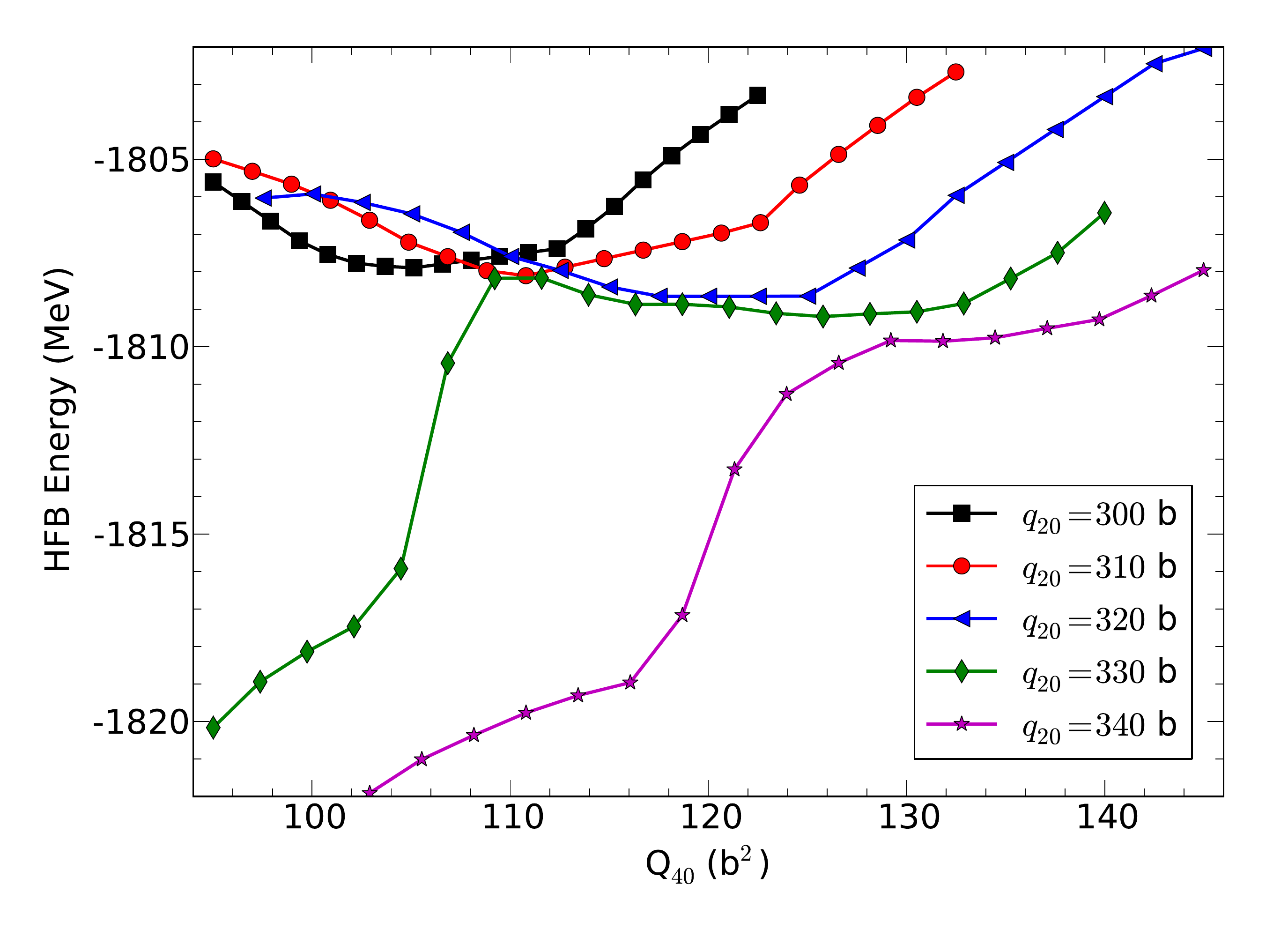}
\caption{(color online) Variation of the total HFB energy as a function of
the hexadecapole moment $q_{40}$ along the least-energy fission pathway in
$^{240}$Pu.
}
\label{fig:PES_1D_cutQ40}
\end{figure}

Because of the risk of discontinuities, we have employed various methods to
ensure that the one-dimensional fission pathway is truly the lowest energy path
connecting the ground-state to the scission point, at least within the
numerical accuracy of the calculations. In particular, we verify {\it a
posteriori} in Fig.~\ref{fig:PES_1D_cutQ40} the correctness of the calculation
in the scission region by showing cross-sections of the energy as a function of
the hexadecapole moment at several points along the path. Together with the
two-dimensional PES of Fig.~\ref{fig:PES_2D_q20q40}, it confirms that our
least-energy fission pathway stays within the fission valley, and, therefore,
corresponds as expected to the hot fission process.


\subsection{Dependence on the Energy Functional}
\label{subsec-skyrmeDependence}

Previous studies carried out with the finite-range Gogny pseudopotential and
the D1S parametrization showed that symmetric fission occurs at very large
values of the quadrupole moment, around $q_{20} \approx 590\, \text{b}$
\cite{(You09)}. We report qualitatively similar results with the Skyrme SkM*
parameterization, although the actual value of the quadrupole moment is
significantly lower, around $q_{20} \approx 550\, \text{b}$. Similarly, the hot
scission point for asymmetric fission is located around
$(q_{30}, q_{40}) \approx (64\,\text{b}^{3/2}, 187\,\text{b}^{2})$ for the
Gogny D1S, while it is
$(q_{30}, q_{40}) \approx (40\,\text{b}^{3/2}, 136\,\text{b}^{2})$ for the SkM*
parameterization. Since we have verified that the one-dimensional fission
pathways reported earlier are truly at the bottom of the fission valley (see
previous section), it is highly unlikely that the differences observed between
D1S and SkM* originate from numerical or algorithmic errors. Instead, they
should be attributed to the intrinsically different deformation properties of
each EDF. In this section, we explore the sensitivity of both the full fission
pathway and the position of the scission configurations on the form of the
energy density used.


\subsubsection{Dependence on the Skryme Energy Density}
\label{subsubsec-skyrme}

\begin{figure}[!ht]
\center
\includegraphics[width=\linewidth]{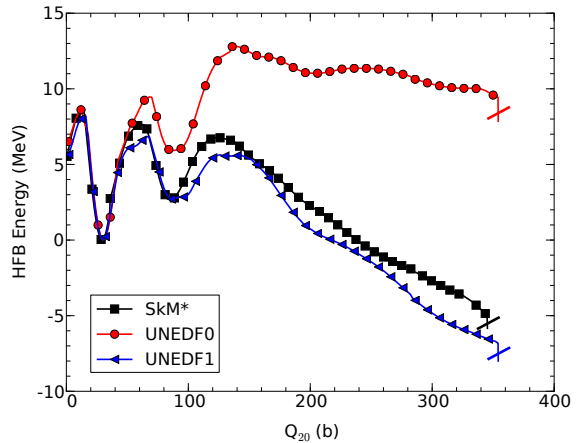}
\caption{(color online) Energy along the least-energy fission pathway in
$^{240}$Pu for three parameterizations of the Skyrme functional, SkM*
\cite{(Bar82)}, UNEDF0 \cite{(Kor10)} and UNEDF1 \cite{(Kor12)}. All curves
are given relative to their ground-state value.
}
\label{fig:skyrme}
\end{figure}

To investigate further the dependence of the scission point on the
parametrization of the energy functional, we have computed the least-energy
fission pathway with the UNEDF0 \cite{(Kor10)} and UNEDF1 functionals
\cite{(Kor12)}. Benchmarks of fission barriers and fission isomer excitation
energies were already reported and discussed in Refs.~\cite{(McD13),(Kor12)}.
Here, we push the calculation up to the scission point and beyond. In this
section, scission is simply defined as the occurrence of a sharp discontinuity
in the PES before which the nucleus is whole ($q_{N} \gg 1$), and after which
it is made of two fragments ($q_{N} \ll 1$). The energy along the least-energy
fission path is shown in Fig.~\ref{fig:skyrme}, and the position of the
scission point is summarized in Table~\ref{table:skyrme}.

\begin{table}[!ht]
\begin{center}
\caption{Approximate position of the scission point in the 
$(q_{20}, q_{30}, q_{40})$ plane for the three parameterizations of the Skyrme 
functionals, SkM*, UNEDF0 and UNEDF1.
}
\begin{ruledtabular}
\begin{tabular}{cccc}
Functional & $\langle \hat{Q}_{20}\rangle$ (b) & $\langle \hat{Q}_{30}\rangle$
(b$^{3/2}$) &
$\langle \hat{Q}_{40}\rangle$ (b$^{2}$) \\
SkM*       & 345 &  43 & 136 \\
UNEDF0     & 354 &  44 & 144 \\
UNEDF1     & 354 &  45 & 146 \\
\end{tabular}
\end{ruledtabular}
\label{table:skyrme}
\end{center}
\end{table}

Interestingly, the position of the scission point is nearly the same for
UNEDF0 and UNEDF1, even though the prescission energy (difference between the
potential energy at the top of the second barrier and at scission) is
remarkably different, with approximately 12.5 MeV for UNEDF1 and only 3.4 MeV
for UNEDF0. These differences in deformation energy are especially striking
since these two functionals give very similar results across a broad range of
nuclear observables including atomic masses, radii, odd-even mass differences,
neutron droplets, etc.. They are most likely caused by the large difference in
the surface-symmetry energy between the two functionals,
$a_{\text{ssym}} = -44 $ MeV for UNEDF0, $a_{\text{ssym}} = -29 $ MeV for
UNEDF1, which decreases significantly surface tension effects \cite{(Nik11)}.

\subsubsection{Dependence on the Pairing Strength}
\label{subsubsec-pairing}

One trademark of the UNEDF family of Skyrme functionals is that the two pairing
strengths of the functional (\ref{eq:pairing}) are fitted simultaneously with
the coupling constants of the Skyrme functional, i.e., the particle-hole and
particle-particle channel of the EDF are treated on the same footing. In
addition, these functionals are used with an approximate formulation of the
Lipkin-Nogami prescription to limit the fluctuations in particle number. The
different fission pathways and scission configurations reported in the previous
section could, therefore, be attributed either to the Skyrme functional itself,
to the pairing channel, or to a complex interplay between the two. In this
section, we briefly analyze the role of pairing correlations alone.

\begin{figure}[!ht]
\center
\includegraphics[width=\linewidth]{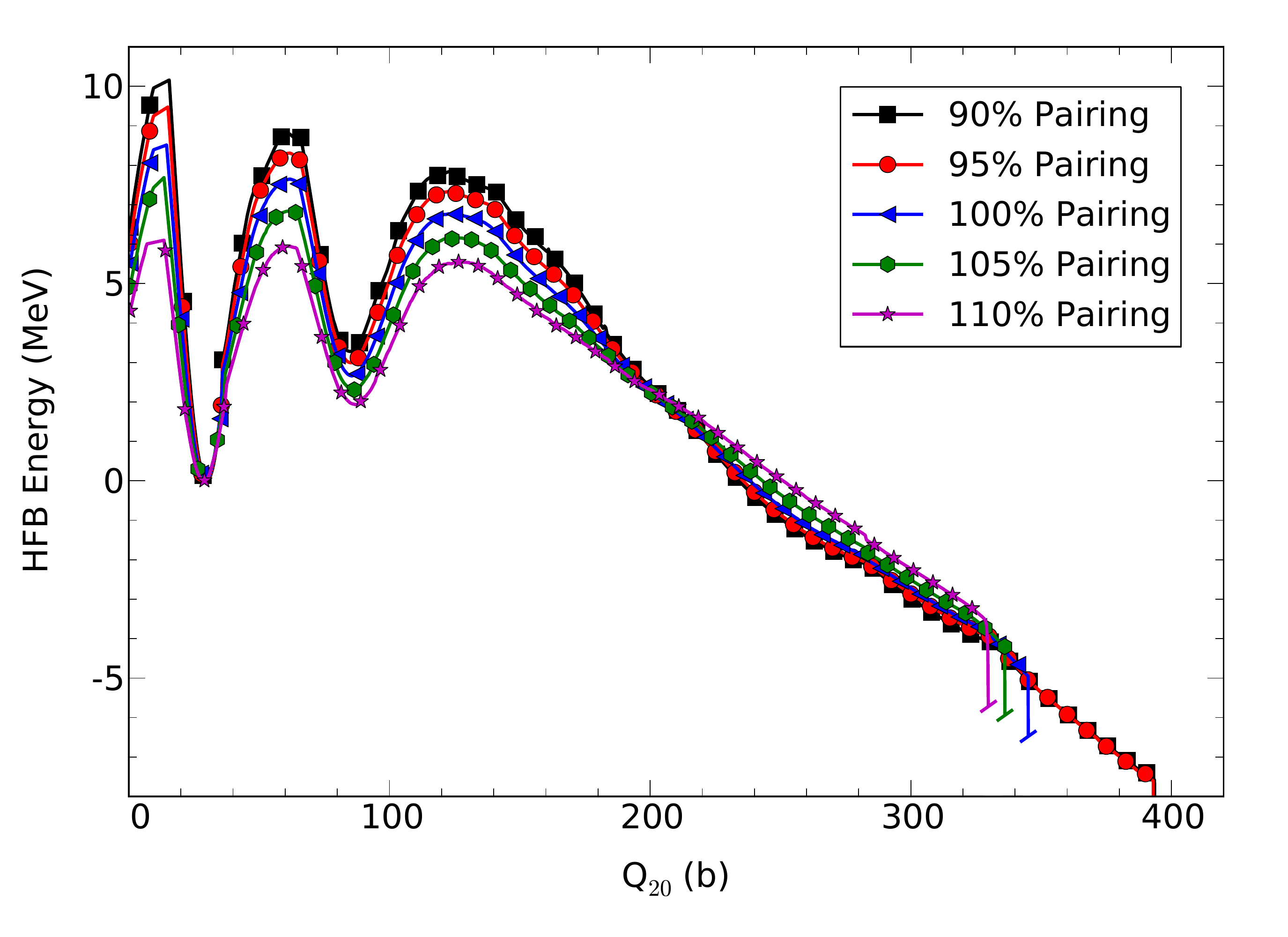}
\caption{(color online) Energy along the least-energy fission pathway in
$^{240}$Pu for five parameterizations of the pairing force and the SkM*
parametrization of the Skyrme functional. All curves are given relative to
their ground-state value.
}
\label{fig:pairing}
\end{figure}

In Fig.\ref{fig:pairing}, we have performed additional calculations of the
fission pathway in $^{240}$Pu by varying both pairing strengths $V_{0}^{(n)}$ and
$V_{0}^{(p)}$ by -10\%, -5\%, +5\% and +10\%. Variations of $\pm5$\% of the
pairing strength leads to variations of about 250 keV of the pairing gaps
computed in the g.s.. This value is often taken as an estimate of the
predictive power of surface-volume pairing interaction combined with Skyrme
functionals to reproduce odd-even mass differences \cite{(Ber09a)}.

The effect of pairing correlations on fission barrier and collective inertia is
well-known, see, e.g., \cite{(Giu13)} and references therein. Less known is the
impact of pairing correlations on the scission point. We find that increasing
pairing decreases the value of the quadrupole moment where scission occurs.
Conversely, decreasing pairing moves the scission configurations to larger
quadrupole moments. The effect is particularly pronounced if pairing
correlations vanish: for pairing strengths decreased by both 5\% and 10\%,
neutron pairing correlations are 0 beyond $q_{20} > 238$ b, resulting in a
shift of the scission point by nearly 50 b compared to the original
calculation. This result suggests that a predictive theory of nuclear fission
based on DFT will require a very accurate description of pairing correlations.


\subsection{Scission Region}
\label{subsec-scission}

By contrast to current theories of spontaneous fission, which rely on the
detailed knowledge of the potential energy surface only in the vicinity of the
ground-state and the two fission barriers, models of induced fission need to
describe the collective space up to, and beyond, the point of scission. Below,
we discuss some of the features of the PES in the scission region for
$^{240}$Pu.


\subsubsection{Triaxiality at and Beyond Scission}
\label{subsubsec-triaxiality}

While the impact of triaxiality on fission barriers has been established for
over forty years, little else is known about the role of this degree of freedom
in the fission process. The additional cost of breaking axial symmetry is
significant, both computationally and physically (loss of the $K$ quantum
number). The purpose of this section is to highlight the role of triaxial
shapes at scission and beyond.

\begin{figure}[!ht]
\center
\includegraphics[width=\linewidth]{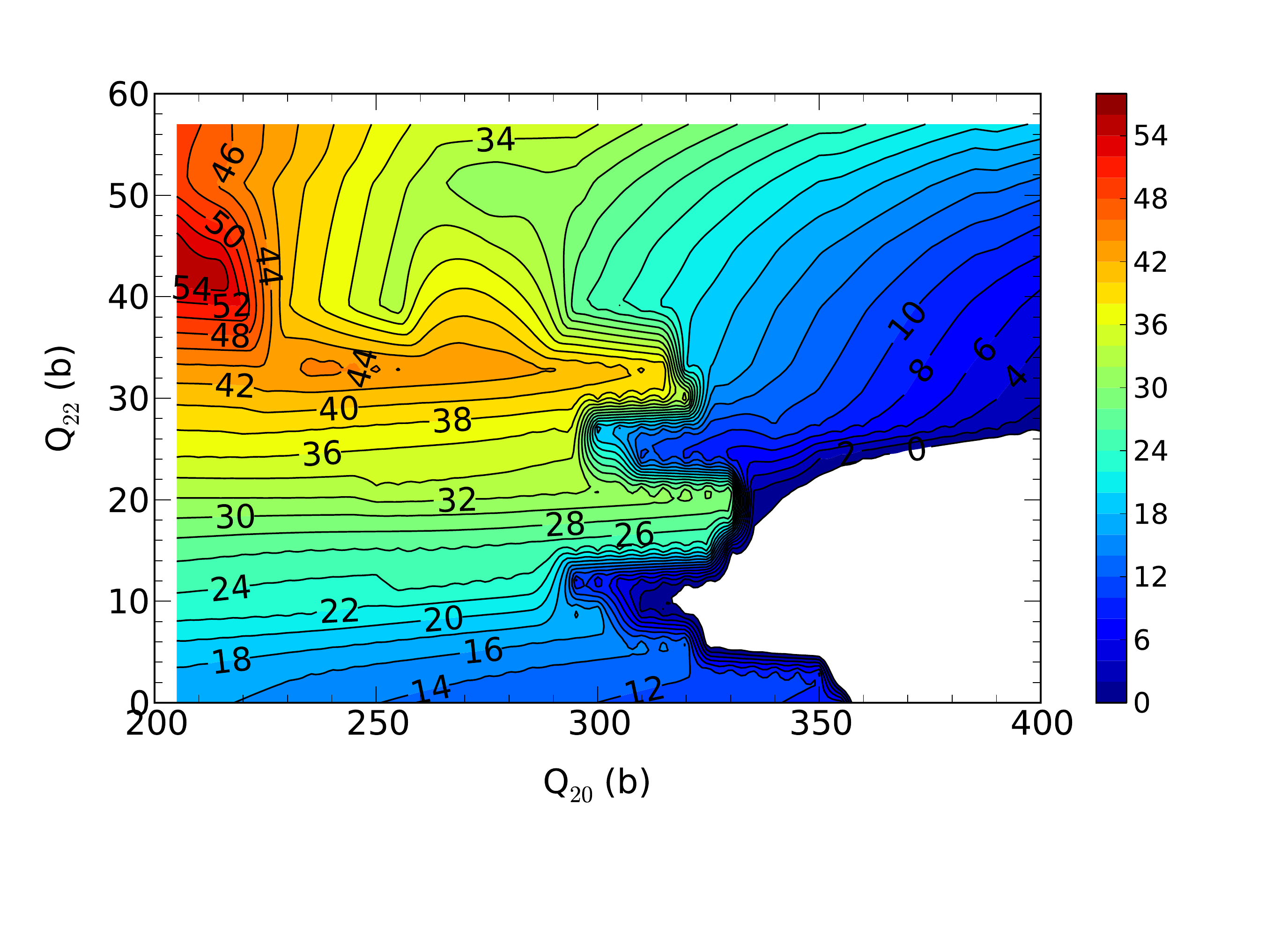}
\caption{(color online) Two-dimensional potential energy surface of $^{240}$Pu
in the $(q_{20},q_{22})$ plane for the SkM* functional around the least-energy
fission pathway. The energy is normalized arbitrarily at -1820 MeV.
}
\label{fig:PES_2D_q20q22_scission}
\end{figure}

We show in Fig.~\ref{fig:PES_2D_q20q22_scission} the potential energy of
$^{240}$Pu for the SkM* functional in the $(q_{20},q_{22})$ plane near
scission. Calculations are based on the least-energy fission pathway of
Fig.~\ref{fig:PES_1D_energy}. For each point in the $(q_{20}, q_{22})$ mesh of
Fig.~\ref{fig:PES_2D_q20q22_scission}, the HFB calculation is initialized with
the nearest HFB solution along the least-energy fission pathway, with the
additional condition that the initial solution satisfies $q_{20} < 300$ b. The
purpose of this last condition is to ensure that the initial guess for the HFB
solution corresponds to a whole nucleus and not two fragments. The resulting
map can be interpreted as a local two-dimensional cross-section in the
$(q_{20}, q_{22})$ along the least-energy fission pathway.

\begin{figure}[!ht]
\center
\includegraphics[width=\linewidth]{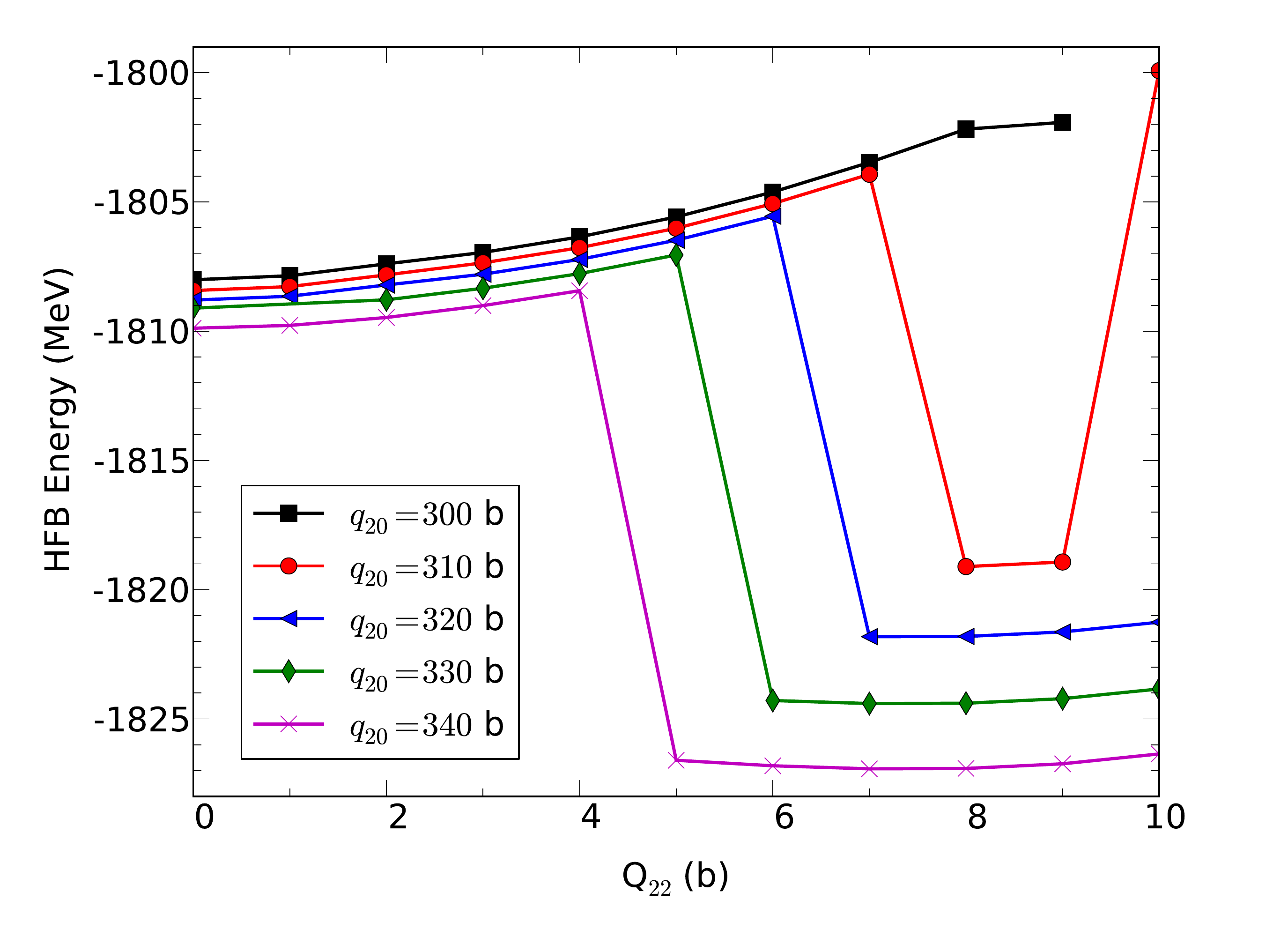}
\caption{(color online) One-dimensional potential energy surface of $^{240}$Pu
along the $q_{22}$ direction for the SkM* functional around the least-energy
fission pathway.
}
\label{fig:PES_1D_q20q22_scission}
\end{figure}

Figure \ref{fig:PES_2D_q20q22_scission} suggests that the least-energy fission
pathway corresponds to a relatively flat valley in the $(q_{20}, q_{22})$
plane. We note that scission has also occurred in the region with
$q_{22} > 40 \,\text{b}$ (with $\gamma \approx 10^{o}$), but the 40 MeV barrier
should in practice hinder this scenario for the range of excitation energy
considered here. We show in Fig.~\ref{fig:PES_1D_q20q22_scission}
one-dimensional cross-sections of the surface for selected values of $q_{20}$
in the range $0 \leq q_{22} \leq 10\,\text{b}$. At $q_{20} = 310\,\text{b}$,
the scission barrier is about 6 MeV high, and only 1.5 MeV high at
$q_{20} = 340\,\text{b}$. Note that the values of $q_{20}$ and $q_{22}$
correspond to very small triaxiality of at most $\gamma \approx 1^{o}$: from a
computational point of view, therefore, there is very little K-admixture in HFB
states. However, even such tiny effects can have a sizable impact on fission
fragment properties as they shift the scission point to lower elongations:
Table~\ref{table:fragment_triax} lists the average proton and neutron number of
the fission fragments at the triaxial scission points. There is a variation of
about 0.5 proton and 1 neutron across this region.

\begin{table}[!ht]
\begin{center}
\caption{Variation of the light (L) and heavy (H) fragment proton and neutron
numbers as a function of triaxiality near the least-energy fission pathway.
}
\begin{ruledtabular}
\begin{tabular}{cccccc}
$q_{20}$ (b) & $q_{22}$ (b) & Z$_{H}$ & N$_{H}$ & Z$_{L}$ & N$_{L}$ \\
310.0        & 7.0          & 53.6   & 84.5   & 40.4   & 61.4   \\
320.0        & 6.0          & 53.7   & 84.8   & 40.3   & 61.2   \\
330.0        & 5.0          & 53.9   & 85.2   & 40.1   & 60.8   \\
340.0        & 4.3          & 54.0   & 85.5   & 40.0   & 60.6   \\
\end{tabular}
\end{ruledtabular}
\label{table:fragment_triax}
\end{center}
\end{table}

The modification of the fission fragment properties induced by triaxiality
should be visible in a dynamical description of fission such as the
time-dependent generator coordinate method \cite{(Gou04),(Gou05)}. The relative
flatness of the collective space in the $(q_{20},q_{22})$ plane should indeed
divert a fraction of the collective flux, which will impact the relative charge
and mass distributions of the fragments. In addition, we may expect a non-zero
dissipation in energy in the transverse collective modes, here characterized by
the $q_{22}$ collective variable, which should reduce the available prescission
energy \cite{(You12b)}.


\subsubsection{Continuous Evolution Across the Scission Point}
\label{subsubsec-scission}

As discussed extensively in Ref.~\cite{(You09)}, an accurate prediction of
fission fragment properties is not possible if the collective space is
restricted to the $(q_{20}, q_{30}, q_{40})$ variables, see also
Sec.~\ref{subsec-fragments} below. Including the triaxial degree of freedom
does not fundamentally alter these conclusions: in such restricted collective
spaces, scission still manifests itself by a sharp discontinuity of the
potential energy surface. Just before this discontinuity, the pre-fragments are
heavily entangled with the consequence that the calculated total kinetic energy
is totally unrealistic; Just after the discontinuity, however, the fragments
are neatly separated but in their ground-state: this is a consequence of the
variational principle behind the HFB approach, and is in contradiction with the
experimental evidence that fission fragments are excited after scission.

\begin{figure}[!ht]
\center
\includegraphics[width=\linewidth]{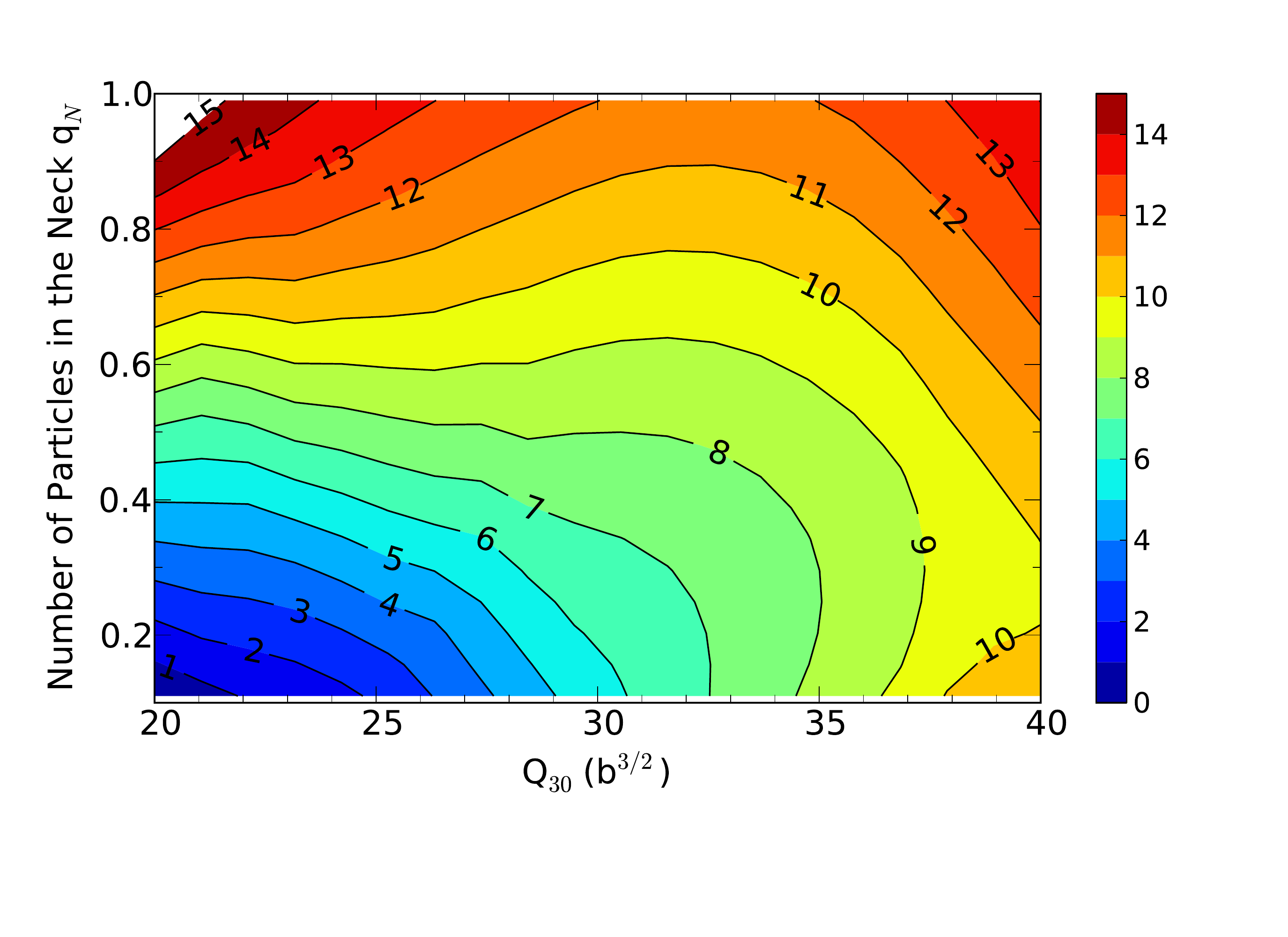}
\caption{(color online) Potential energy surface in the $(q_{30},q_{N})$ plane
just before scission for the SkM* functional. The axial quadrupole moment
$\hat{Q}_{20}$ is fixed at $q_{20} = 345$ b, the triaxial quadrupole
$\hat{Q}_{22}$ and hexadecapole moments $\hat{Q}_{40}$ are unconstrained. The
energy is normalized arbitrarily at -1827 MeV.
}
\label{fig:PES_2D_q30qN}
\end{figure}

It is, in fact, quite simple to introduce additional collective variables that
will transform the discontinuity at scission into a continuous pathway. Among
the possible choices, a constraint $\hat{Q}_{N}$ on the density of particles in
the neck between the two pre-fragments has often been used, both in the context
of spontaneous fission \cite{(War02)}, and induced fission \cite{(You09)}. We
show in Fig.~\ref{fig:PES_2D_q30qN} a close-up of the local potential energy
surface of $^{240}$Pu near the scission point for the SkM* functional. The
axial quadrupole moment is fixed at $q_{20} = 345$ b while the triaxial
quadrupole $\hat{Q}_{22}$ and hexadecapole moments $\hat{Q}_{40}$ are
unconstrained. Only the range $[0,1]$ of $q_{N}$ is represented, as it is in
this area that the scission process seems to take place (see discussion in
Sec.~\ref{subsec-viz}). At $q_{N} = 1$, the least-energy fission pathway
emerges at $q_{30} \approx 30$ b$^{3/2}$. It broadens up to form a wide
``estuary'' in the $(q_{30},q_{N})$ subspace: the energy surface is very
shallow across a large range of octupole moments. This should manifest itself
by a sizable broadening of the yields.

\begin{figure}[!ht]
\center
\includegraphics[width=\linewidth]{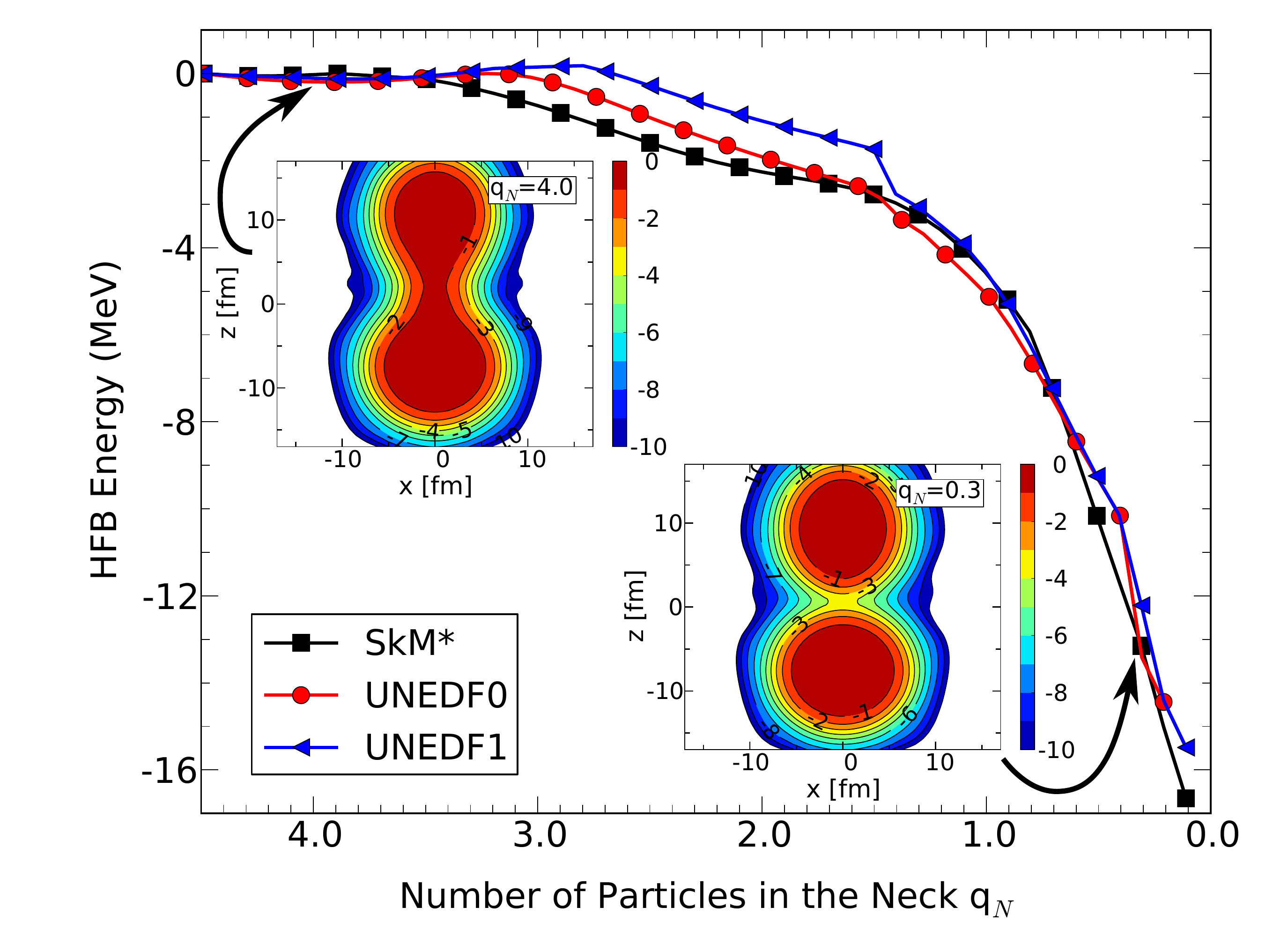}
\caption{(color online) Total energy as a function of the density of particles
in the neck $q_{N}$ along the least-energy fission pathway for the SkM*, UNEDF0
and UNEDF1 functionals. All curves are normalized relative to their respective
values at $q_{N}=4.5$. Inset contour plots show the density profile at
$q_{N} = 4.0$ and $q_{N}=0.3$.
}
\label{fig:PES_1D_qN}
\end{figure}

In order to better visualize the variations in energy when following this
continuous path, we show in Fig.~\ref{fig:PES_1D_qN} the one-dimensional
profile of the total energy as a function of $q_{N}$ for the three functionals
used in this work. For each curve, the value of the axial quadrupole moment is
fixed at the value just before scission as listed in Table~\ref{table:skyrme},
and calculations are performed with a constraint on $\hat{Q}_{N}$. All other
multipole moments are unconstrained. The curves are normalized at the value of
$q_{N} = 4.5$. It is worth noticing that the energy gain along this extra
dimension in the collective space is very similar for all three functionals,
even though the potential energy surface in the $q_{20}$ direction can be
dramatically different, see Fig.~\ref{fig:PES_1D_energy}. On average,
variations of $q_{N}$ lower the energy by up to 12-15 MeV.

Most importantly, this new degree of freedom provides a mechanism to pass
{\em continuously} from a single whole nucleus to two distinct fragments. The
$q_{N}$ degree of freedom can, therefore, be viewed as a kind of control
parameter. It can be used in several ways. The scission configurations can be
chosen along the $q_{N}$ axis arbitrarily, on the sole basis of
phenomenological comparisons with experimental data, e.g. on charge and mass
distribution of fission fragments. Alternatively, additional criteria can be
invoked to pin down the scission configuration at a given value of $q_{N}$, or
in a given interval of $q_{N}$ values. This is the approach that we chose and
that we discuss in more details in the next section.


\section{Nuclear Scission and Fission Fragment Properties}
\label{sec-scission}

The purpose of any theory of induced fission is to predict fragment properties
such as charge and mass distribution, kinetic energy, excitation energy of each
fragment, fission spectrum, etc., as these correspond to measurable quantities.
In the nuclear DFT approach, computing these properties require introducing
scission configurations in the compound nucleus. After a brief
historical reminder, we present below the methods that we used to define the
scission configurations, as well as its application in the calculation of
fission fragment properties for the least-energy fission pathway of $^{240}$Pu.


\subsection{On the Definition of Scission}
\label{subsec-history}

The concept of a scission point has its origin in the liquid drop (LD) picture
of the nucleus and reflects the fact that for very large deformations, the LD
potential energy is a multi-valued function of the deformation parameters
\cite{(Bra72),(Nix65),(Str63)}. These multivalues generate discontinuities in
potential energy landscapes, which are still widely used as a criterion to
define the scission configurations \cite{(You09),(Sta09),(Dub08),(Gou04)}.
However, as we have recalled in the previous section, these discontinuities are
entirely spurious since locally enlarging the collective space can easily
restore the continuity of the full PES \cite{(Dub12)}. In addition, continuous
PES give additional flexibility to define the scission configurations and
improve the predictive power of the theory.

As mentioned above, rather than use the $q_{N}$ degree of freedom as a simple
control parameter that we could tune to data, we would like to find general
criteria, based either on mathematics and/or on physics, to define the scission
configurations, and let the theory take care of the comparison with the data
without further empirical adjustments. The simplest criterion one could invoke
to define scission is to set a minimum value for the size of the neck,
$q_{N}^{\text{(min)}}$, below which one assumes the neck is small enough that
the two fragments can be considered fully formed \cite{(You09),(Dub08)}. Such
an approach has the advantage to be easy to automate, but the choice of
$q_{N}^{\text{(min)}}$ remains entirely arbitrary. A possible extension would
be to set up a {\it range} in $q_{N}$ values, say
$I_{q} = [q_{N}^{\text{(min)}}, q_{N}^{\text{(max)}}]$, where scission
configurations are chosen, and use the boundaries of this interval as estimates
of theoretical errors. Ideally, this interval should be as narrow as possible.
Since at scission one connected nucleus becomes two un-connected fragments,
tools based on detecting connectivity features in datasets (here the nuclear
density) should be applicable, and may help in making the determination of
$I_{q}$ less arbitrary. We will explore this option in details in
Sec.\ref{subsec-viz}.

For the sake of completeness, we also mention an alternative strategy to
identify scission configurations. It was recognized early on that the
competition between the repulsive Coulomb and the attractive nuclear force may
induce the scission of the nucleus even when there still remains a sizable neck
between the two nascent fragments: the ratio of the Coulomb energy over the
nuclear interaction energy can, therefore, provide a complementary, dynamical,
criterion for scission \cite{(Dav76),(Bon07a)}. Recently, a similar approach
was formalized in the context of nuclear density functional theory
\cite{(You11)}. Two major ingredients are required: (i) the calculation of the
Coulomb and nuclear interaction energy, (ii) a procedure to "localize" the
fragments, since both the Coulomb and nuclear interaction energy are
representation-dependent, see Sec.\ref{subsec-localization} below. These
approaches can of course be combined with the method we describe below.


\subsection{Topological Identification of the Scission Point}
\label{subsec-viz}

As mentioned above, one possible way to define scission configurations is to
use computational tools that detect connectivity features in datasets. The
problem we are posing is thus the following: given a set of neutron and proton
densities $\{ \rho_{n}(\gras{r}), \rho_{p}(\gras{r}) \}_{q_{N}}$ computed at
each point along some trajectory -- here parametrized by the expectation value
$q_{N}$ of the neck operator, is it possible to identify changes in the
densities that could be interpreted as the transition from a single nucleus to
two fragments? It is one instance of a wider problem in computational science,
that of identifying, within data, combinatorial changes that are assumed to be
markers for a physical phenomena. In fact, it is possible to characterize such
combinatorial changes using a vocabulary that is {\it independent} of
assumptions about numerical thresholds in the physical system. In place of
geometric properties, our approach draws on the mathematics of topology to
derive global properties (invariants) of space. Topological analysis of
datasets results in structural abstractions that can be used to answer
questions of whether two spaces have fundamentally the same shape, and to
articulate the types of differences that appear. In the physical sciences,
topological feature analysis is well known -- in the study of flow (e.g.
vortices, separatrices) \cite{(Hel89),(Tri00),(The03)}, and in the analysis of
scalar fields (e.g. critical points) \cite{(Nat06),(Lan06),(Day09)}. Its main
advantage is that it moves identification of phenomena from assessment against
empirical, possibly subjective, thresholds into binary decisions based on
change to the discrete structures that express fundamental topological
properties. This leaves two questions: which (if any) topological change in
data correlate with the physical phenomena of interest, and can topological
structure be computed ``effectively''?

\begin{figure}[hbt]
\centering
\includegraphics[width=.48\textwidth]{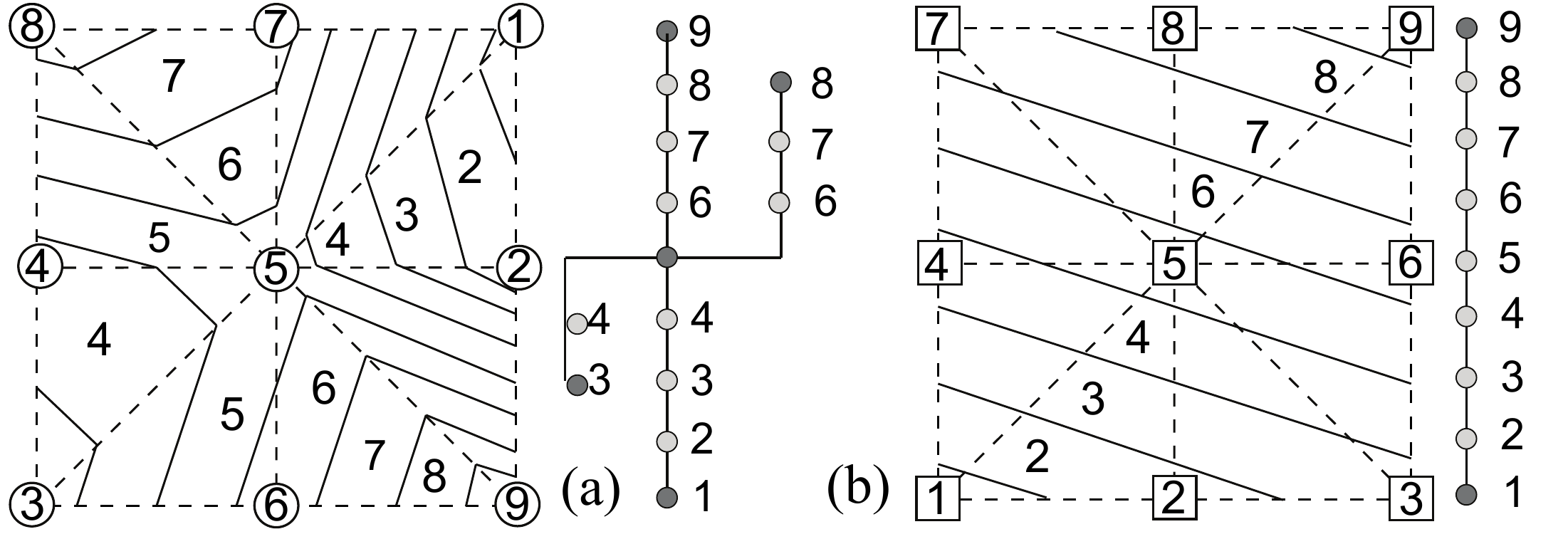}
\caption{Two small functions $f(x,y)$ and $g(x,y)$ on a triangular mesh (dashed
lines) shown as contour plots and their respective contour trees. Circled (a)
and squared (b) points represent the values of the function on the mesh. }
\label{fig:contourTree}
\end{figure}

Recent work in scientific visualization and computational topology has shown
how to analyze features in functions of the form $f: \mathbb{R}^3 \rightarrow
\mathbb{R}$, such as the local nuclear density $\rho(\gras{r})$. In these functions,
the connectivity of isovalued contours can be analyzed using the {\it contour
tree}~\cite{CSV10}, which captures the relationships of all possible contours in
a data set. Fig.~\ref{fig:contourTree} gives a pedagogical illustration of the
technique: maxima and minima of the contour map are leaves of the tree, while
critical points (saddles) are interior nodes. Moreover, one characterization of
critical points is that they are the highest points at which two peaks are
connected. As a result, the critical points naturally define features
corresponding to branches of the tree. Subsequent work showed that these
features can then be tracked over time (or any other relevant parameter)
\cite{BWPDB10}.

\begin{figure}[hbt]
\centering
\includegraphics[width=.48\textwidth]{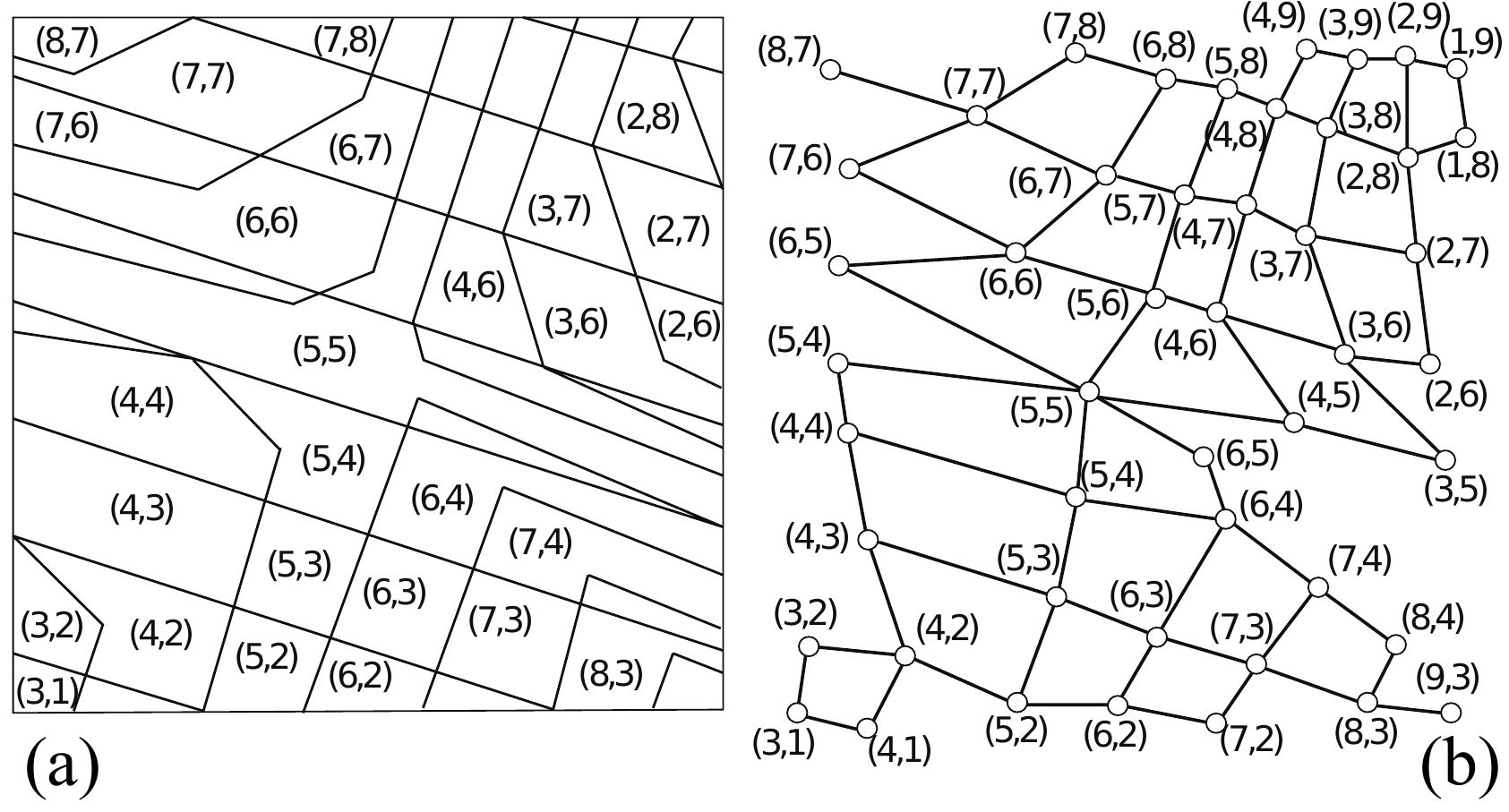}
\caption{Joint contour slabs found by intersecting the slabs of functions $f$
and $g$ of Fig.~\ref{fig:contourTree} (a); Joint Contour Net obtained by
analyzing the connectivity of the slabs, see text for details (b).}
\label{fig:smallJCN}
\end{figure}

While this approach works for single-valued functions, it needs modification
for bivariate functions of the form
$(f,g): \mathbb{R}^3 \rightarrow \mathbb{R}^{2}$. For such a function, contours
do not naturally divide it into features, and a generalization of the contour
tree is required, as shown in Fig.~\ref{fig:smallJCN}. Here, the domain is
divided along contours of both $f$ and $g$, resulting in a set of regions, or
``slabs'', as shown. To understand how the abstract graph of panel (b) is
obtained from the original contours of panel (a), consider for instance the
lower left corner of panel (a): the slab marked (3,2) is adjacent to the slabs
(3,1) and (4,2). Hence, the node (3,2) is connected to the two nodes (4,2) and
(3,1). Systematically analyzing the connectivity of these slabs thus gives the
{\it Joint Contour Net} (JCN) shown in Panel (b) of Fig.~\ref{fig:smallJCN}: an
abstract representation of the joint variation of $f$ and $g$~\cite{CD13}.

The case of nuclear fission lends itself perfectly to such an analysis. Indeed,
within nuclear DFT the nucleus is entirely characterized by the neutron
($\rho_{n}$) and proton ($\rho_{p}$) densities, which will play the role of the
two distinct yet correlated scalar fields $f$ and $g$ in the example of figures
\ref{fig:contourTree}-\ref{fig:smallJCN}. Any special feature of the JCN graph
associated with the bivariate function
$(\rho_{p},\rho_{n}): \mathbb{R}^3 \rightarrow \mathbb{R}^2$ could therefore,
in principle, be given a physical interpretation. In fact, we have recently
shown that the sudden division of the compound nucleus in two separate
fragments at the discontinuity of one-dimensional fission pathways $E(q_{20})$
is clearly associated with a fork in the JCN \cite{(DCK12)}. Here, we extend
the method to the more difficult problem of detecting features along a
{\it continuous} fission pathway characterized by the $q_{N}$ constraint.

\begin{figure}[hbt]
\centering
\includegraphics[width=.45\textwidth]{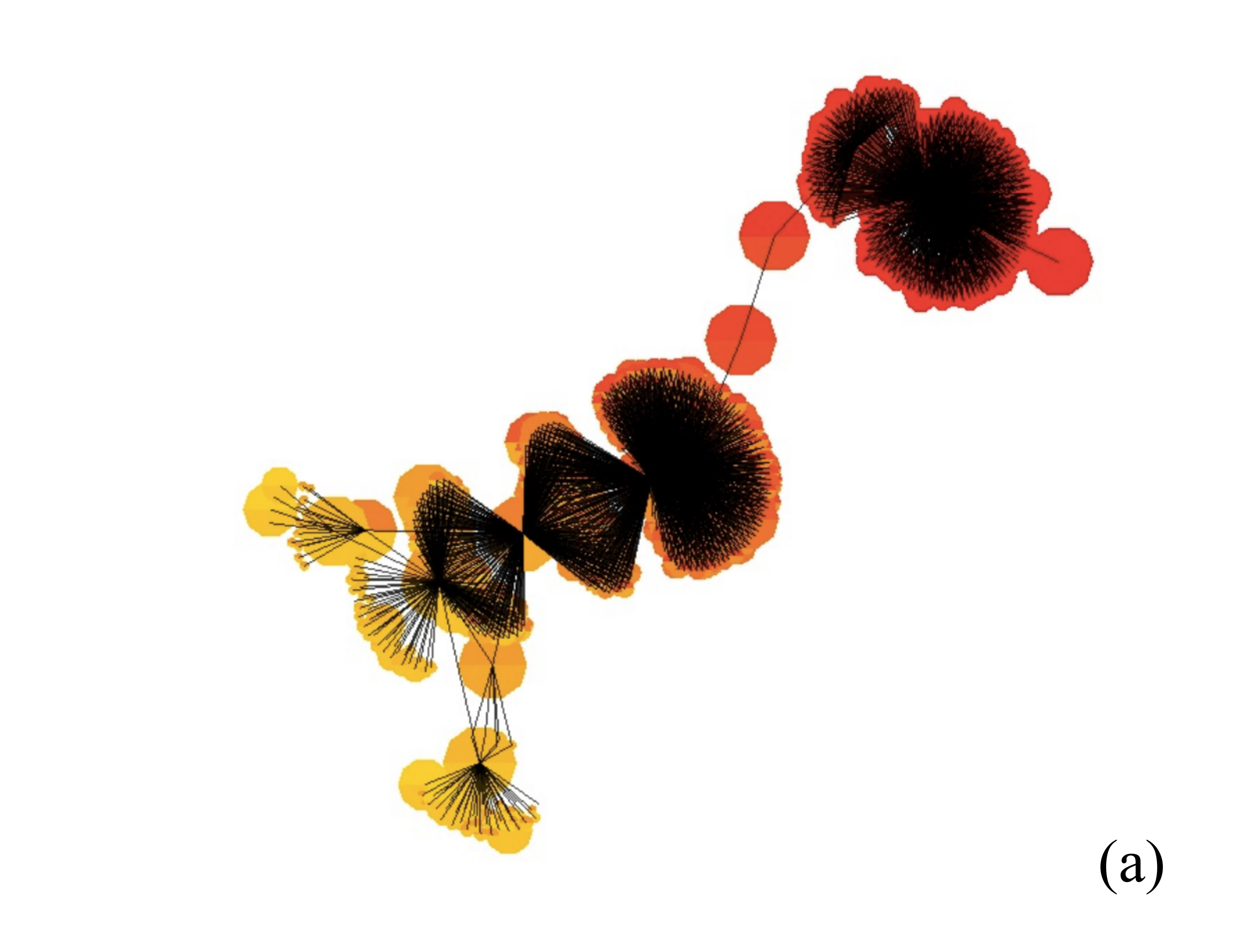}
\includegraphics[width=.45\textwidth]{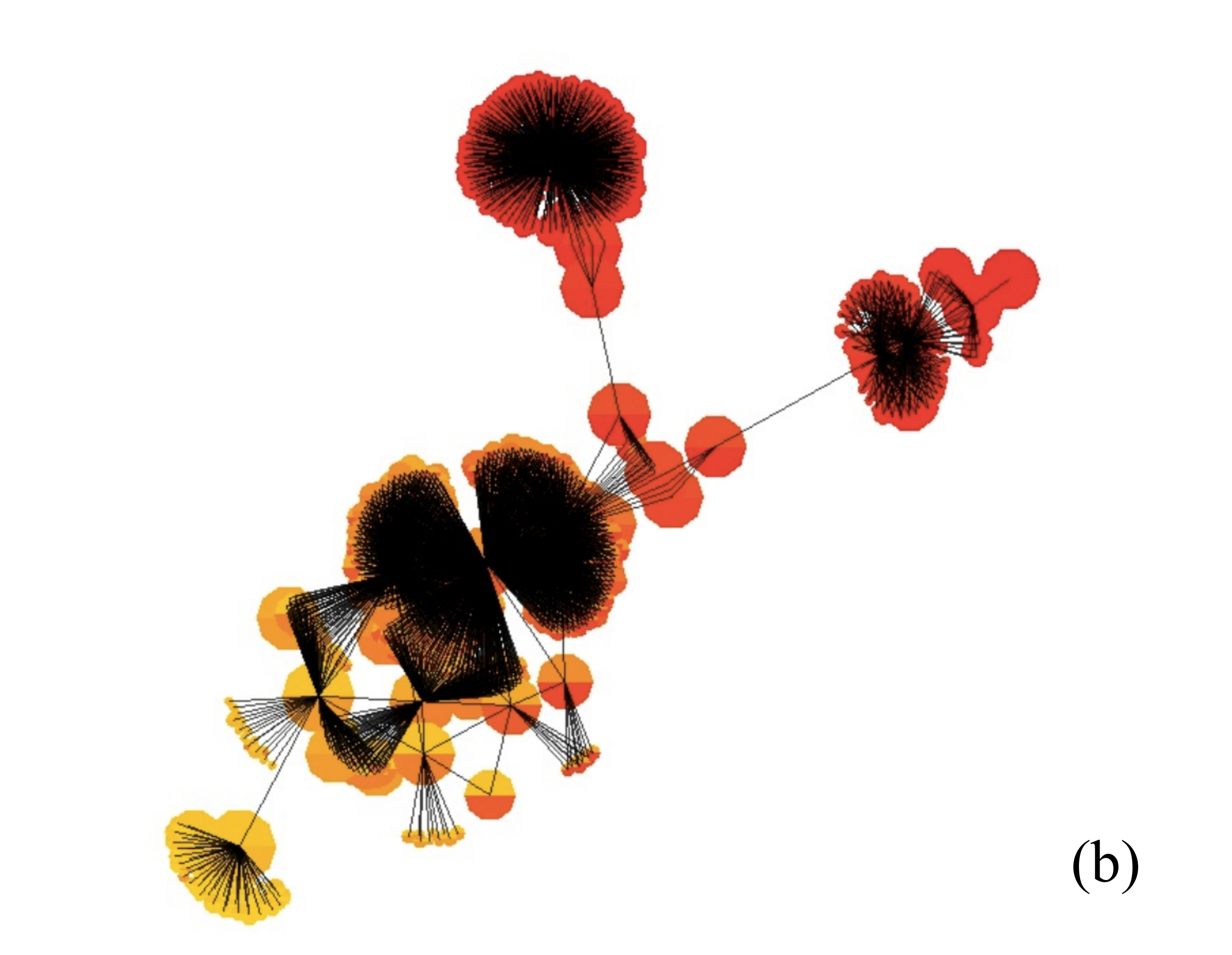}
\caption{(color online) JCN graphs near the scission for $^{240}$Pu at
$q_{N} = 4.0$ (a) and at $q_{N} = 0.1$ (b). The principal feature visible is
that the single branch for high isovalues of the densities (upper right side of
top figure) at $q_{N} = 4.0$ has forked into two distinct high isovalues
branches (upper right side of bottom figure) at $q_{N}=0.1$, each branch
featuring starbursts.}
\label{fig:JCNScission}
\end{figure}

Fig.~\ref{fig:JCNScission} illustrates the application of the JCN method to the
detection of scission in $^{240}$Pu. The contour nets are extracted from the
densities of $^{240}$Pu at the two values $q_{N} = 0.1$ and $q_{N} = 4.0$, see
Fig.~\ref{fig:PES_1D_qN}. The principal visual features of the JCN are forks
and circular structures, which we named starbursts:
\begin{itemize}
\item As recalled above, a fork at the high-density end of the JCN (red, upper
right part of each graph) shows the presence of two distinct features meeting
at a critical point, rather than a single peak, i.e. two topologically distinct
regions of space \cite{(DCK12)}. Here, we interpret the first occurrence of
such a fork at high density values as the {\it precursor} to scission, marking
the upper bound $q^{\text{(max)}}$ of the interval $I_{q}$ defining scission.
\item Subsequent development of the ``starburst'' in each branch suggests that
these two regions acquire independent internal structure. That is, the range
and variation in proton and neutron field density levels in the two distinct
regions is commensurate with that present in the nucleus before the appearance
of branching. Therefore, we interpret the first occurrence of such starbursts
as the signal that the nucleus has split into two {\it well-formed fragments},
which defines the lower bound $q^{\text{(min)}}$ of $I_{q}$.
\end{itemize}
While visual interpretation of the features of the JCN graph relies on
judgements based on calibration experiments, the underlying graph is
mathematically well-defined and its construction is topologically rigorous. The
only input to analysis is the spatial representation of the neutron and proton
densities, and the single output is an estimate of the interval $I_{q}$ where
scission occurs. When scanning the entire range in $q_{N}$ value from 0 to 4.5,
we have found that the interval $I_{q}$ was $I_{q} = [0.2, 2.6]$ for SkM* and
UNEDF0, and $I_{q} = [0.2, 2.2]$ for UNEDF1.

For applications to nuclear fission, Joint Contour Net analysis depends
principally on the level at which the density values are quantized into slabs.
Initial work showed that analysis can detect scission at different levels of
quantization, with finer levels of quantization narrowing the candidate
scission point to a smaller number of sites \cite{(DCK12)}. Beyond a certain
limit no further narrowing was observed, suggesting that the analysis is then
constrained by the data, that is, independent of the quantization level.


\subsection{Fission Fragment Identification}
\label{subsec-identification}

Topological methods such as the JCN can automate the identification of a 
putative scission region in the collective space. In order to compute fission 
fragment properties within this region, the density matrix and pairing tensor 
of each of the fragments must be determined. We start from the set of 
quasi-particles for the compound nucleus defined by the Bogoliubov matrices $U$ 
and $V$. The coordinate space representation of the full one-body density
matrix (in coordinate$\otimes$spin space) reads
\begin{equation}
\rho(\gras{r}\sigma,\gras{r}'\sigma') = \sum_{ij} \rho_{ij}
\phi_{i}(\gras{r}\sigma)\phi_{j}^{*}(\gras{r}'\sigma'),
\end{equation}
with $\phi_{i}(\gras{r}\sigma)$ the basis functions, and
$\rho_{ij} = \sum_{ij} V_{i\mu}^{*}V_{j\mu}$ the configuration space
representation of the density matrix. We can introduce a quasiparticle (q.p.)
density $\rho_{\mu}(\gras{r}\sigma,\gras{r}'\sigma')$ by
\begin{equation}
\rho_{\mu}(\gras{r}\sigma,\gras{r}'\sigma') = \sum_{ij} V_{i\mu}^{*}V_{j\mu}\;
\phi_{i}(\gras{r}\sigma)\phi_{j}^{*}(\gras{r}'\sigma'),
\end{equation}
such that the occupation $N_{\mu}$ of a single quasi-particle $\mu$ is simply
\begin{equation}
N_{\mu} = \sum_{\sigma} \int d^{3}\gras{r}\;
\rho_{\mu}(\gras{r}\sigma,\gras{r}\sigma).
\label{eq:occupation}
\end{equation}
Since the basis $\{ \phi_{i} \}$ is orthonormal, this reduces to the well-known
expression $N_{\mu} = \sum_{ij} V_{i\mu}^{*}V_{j\mu}$, with the total number of
particles defined as $N = \sum_{\mu} N_{\mu}$. Let us assume the neck is
located along the $z$-axis of the intrinsic reference frame, and thus has the
coordinates $\gras{r}_{\text{neck}} = (0,0, z_{N})$. We can then define the
occupation of the q.p. $\mu$ in the fragment (1) as
\begin{equation}
N_{1,\mu} = \sum_{ij} V_{i\mu}^{*}V_{j\mu} d_{ij}(z_{N}),
\label{eq:occupation1}
\end{equation}
where
\begin{equation}
d_{ij}(z) = \sum_{\sigma}
\int_{-\infty}^{+\infty} dx\int_{-\infty}^{+\infty} dy \int_{-\infty}^{z}dz\;
\phi_{i}(\gras{r}\sigma)\phi_{j}^{*}(\gras{r}\sigma).
\end{equation}
The occupation of the q.p. in the fragment (2) is simply
$N_{2,\mu} = N_{\mu} - N_{1,\mu}$. We then assign the q.p. $\mu$ to fragment
(1) if $N_{1,\mu} \geq 0.5 N_{\mu}$, and to fragment (2) if
$N_{1,\mu} <0.5 N_{\mu}$. In this way, the full set of q.p. is partitioned in 
two subsets, each corresponding to one of the fragments.

These two sets of q.p. allow us to build the analogs of the density matrix and 
the pairing tensor for the fragments. In coordinate$\otimes$spin space, we will 
thus define
\begin{eqnarray}
\rho_{\text{f}}(\gras{r}\sigma,\gras{r}'\sigma') 
=
\sum_{\mu\in(\text{f})}  \sum_{ij} V_{i\mu}^{*}V_{j\mu}\;
\phi_{i}(\gras{r}\sigma)\phi_{j}^{*}(\gras{r}'\sigma'), \medskip\\
\kappa_{\text{f}}(\gras{r}\sigma,\gras{r}'\sigma') 
=
\sum_{\mu\in(\text{f})}  \sum_{ij} V_{i\mu}^{*}U_{j\mu}\;
\phi_{i}(\gras{r}\sigma)\phi_{j}^{*}(\gras{r}'\sigma'),
\label{eq:density_fragment}
\end{eqnarray}
with $\text{f}=1,2$ labeling the fragment. Note that, by constrast to the full 
density matrix of the compound nucleus $\rho$, the objects $\rho^{(\text{f})}$ 
are {\em not} one-body densities in the strict mathematical sense. In particular, 
they are not projectors in Fock space, 
$\rho^{(\text{f})2} \neq \rho^{(\text{f})}$. Also, the usual relations 
$\rho^{2} + \kappa\kappa^{\dagger} = 0$ are not necessarily satisfied for 
$\rho^{(\text{f})}$ and $\kappa^{(\text{f})}$. We should therefore refer to 
these objects as pseudodensities, to emphasize their empirical nature. The 
diagonal component of these pseudodensities (in 
coordinate$\otimes$spin space) $\rho^{(1)}(\gras{r})$,
$\rho^{(2)}(\gras{r})$, $\kappa^{(1)}(\gras{r})$ and $\kappa^{(2)}(\gras{r})$ 
for each fragment can be obtained as usual, for example,
\begin{equation}
\rho^{(\text{f})}(\gras{r})
= 
\sum_{\sigma\sigma'} 
\rho^{(\text{f})}(\gras{r}\sigma,\gras{r}\sigma')
\end{equation}
define the local pseudodensity in fragment ``f''. Similarly, one can define the 
analog of the kinetic energy density, $\tau^{(\text{f})}$, and the spin current 
tensor, $J_{\mu\nu}^{(\text{f})}$, for each fragment, as well as their time-odd 
counterparts 
\cite{(Ben03),(Eng75)}.

After the pseudodensity matrix and pairing pseudotensor of each fragment have 
been defined, all fragment energies and interaction energies can be computed in 
a straightforward manner at the HFB approximation. The Coulomb interaction energy 
between the fragments is
\begin{equation}
E_{\text{Cou}} = 
E_{\text{Cou}}^{1\rightarrow 2} + E_{\text{Cou}}^{2\rightarrow 1}.
\label{eq:coulomb_energy}
\end{equation}
For both the direct and exchange term, 
$E_{\text{Cou}}^{1\rightarrow 2} = E_{\text{Cou}}^{2\rightarrow 1}$, hence we find
\begin{equation}
E_{\text{Cou}}^{(\text{dir})} 
= 
2e^{2} \int d^{3}\gras{r}\int d^{3}\gras{r}'
\frac{\rho^{(1)}(\gras{r})\rho^{(2)}(\gras{r}')}{|\gras{r} - \gras{r}'|},
\label{eq:coulomb_energy_dir}
\end{equation}
while the (attractive) exchange Coulomb interaction energy is defined by 
\begin{equation}
E_{\text{Cou}}^{(\text{exc})} 
= 
2e^{2} \int d^{3}\gras{r}\int d^{3}\gras{r}'
\frac{\rho^{(1)}(\gras{r},\gras{r}')\rho^{(2)}(\gras{r}',\gras{r})}{|\gras{r} - \gras{r}'|}.
\label{eq:coulomb_energy_exc}
\end{equation}
In these expressions, $\rho^{(1)}$ is the pseudodensity in fragment (1), 
$\rho^{(2)}$ the isoscalar density in fragment (2), and 
$e^{2} = \hbar c /\alpha $ is in MeV.fm. In our calculations, the direct 
Coulomb energy was computed using the Green function method as in 
Ref.~\cite{(Dob97c)} while we used the Slater approximation for the exchange 
part. The (attractive) nuclear interaction energy, which, in our case, is the 
Skyrme interaction energy, is similarly given by
\begin{equation}
E_{\text{nuc}}^{\text{Skyrme}} = 
E_{\text{nuc}}^{1\rightarrow 2} + E_{\text{nuc}}^{2\rightarrow 1},
\label{eq:skyrme_energy}
\end{equation}
and
\begin{multline}
E_{\text{nuc}}^{1\rightarrow 2}
=
\sum_{t=0,1} \int d^{3}\gras{r}
\left\{
C_{t}^{\rho}       \rho_{t}^{(1)}\rho_{t}^{(2)} +
C_{t}^{\Delta\rho} \rho_{t}^{(1)}\Delta\rho_{t}^{(2)} + \right. \\
\left.
C_{t}^{\tau}           \rho_{t}^{(1)}\tau_{t}^{(2)} +
C_{t}^{J}\sum_{\mu\nu} J_{\mu\nu,t}^{(1)}J_{\mu\nu,t}^{(2)} +
C_{t}^{\nabla J}       \rho_{t}^{(1)}\gras{\nabla}\cdot\gras{J}_{t}^{(2)} 
\right\}.
\label{eq:interaction_energy}
\end{multline}
Permute indices 1 and 2 to obtain the second term in 
Eq.(\ref{eq:skyrme_energy}). Note that, contrary to the Coulomb energy, it is 
not symmetric under permutation of the fragments, i.e., 
$E_{\text{nuc}}^{1\rightarrow 2} \neq E_{\text{nuc}}^{2\rightarrow 1}$. 
Because of the zero-range of the Skyrme force, Eq.(\ref{eq:interaction_energy}) 
contains both direct and exchange contributions.


\subsection{Quantum Localization}
\label{subsec-localization}

In this section, we expand on the quantum localization method first introduced
by Younes and Gogny in Ref.~\cite{(You11)}. A consequence of the quantum 
mechanical nature of the system is that the coordinate representations 
$\rho^{(1)}(\gras{r})$ and $\rho^{(2)}(\gras{r})$ of the local pseudodensities 
of each fragment near scission are not clearly localized within their respective 
fragment: the pseudodensity $\rho^{(1)}(\gras{r})$ has a tail that extends 
significantly into fragment (2) and vice-versa, see  Fig.~\ref{fig:density_loc}. 
In the HFB theory, this delocalization of the density can be traced back to the 
individual quasi-particles, and can be captured by the following indicator
\begin{equation}
\ell_{\mu} = \frac{| N_{1,\mu} - N_{2,\mu} |}{N_{\mu}},
\end{equation}
with $N_{\mu}$ defined by Eq.~(\ref{eq:occupation}) and $N_{1,\mu}, N_{2,\mu}$
by Eq.~(\ref{eq:occupation1}). If $\ell_{\mu}=0$, the q.p. $\mu$ is fully
delocalized, if $\ell_{\mu}=1$ it is fully localized either in the left or in
the right fragment. The tails in the pseudodensities are produced by the
contributions from the delocalized q.p. states with relatively large occupation
and $0 \leq \ell_{\mu} \ll 1$.

The larger the overlap between $\rho^{(1)}(\gras{r})$ and 
$\rho^{(2)}(\gras{r})$ is, the larger (in absolute value) the Coulomb and 
nuclear interaction energy, and the lower the fragment intrinsic energies, 
i.e., the higher the excitation energy of the fragments, since, in the HFB 
approximation,
\begin{equation}
E_{\text{tot}}[\rho^{(1)}+\rho^{(2)}]
=
E_{1}[\rho^{(1)}] + E_{2}[\rho^{(2)}] + E_{\text{int}},
\end{equation}
with $E_{\text{int}} = E_{\text{Cou}}^{\text{dir}} + 
E_{\text{Cou}}^{\text{exc}} + E_{\text{nuc}}$. Consequently, we find that both 
the fission fragment properties (total excitation energy, deformation, etc.) 
and the total kinetic energy of the accelerated fragments depend on the overlap 
between $\rho^{(1)}(\gras{r})$ and $\rho^{(2)}(\gras{r})$; Qualitatively, the 
two fragments are entangled.

This entanglement poses a conceptual problem when comparing theoretical
predictions with experimental data on fission fragment properties. Indeed, it
is well-known that the generalized HFB density $\mathcal{R}$ associated with a
given set of q.p. operators $(\beta_{\mu},\beta_{\mu}^{\dagger})$ is invariant
under any unitary transformation of these operators, see e.g.
Refs.~\cite{(Bla85),(Man75)}. While all global observables such as energy,
angular momentum, etc., are invariant, local properties associated with any
subset of the q.p. states may not be. In other words, the energies
$E_{1}[\rho^{(1)}]$, $E_{2}[\rho^{(2)}]$ and $E_{\text{int}}$ are
{\it representation-dependent}: any unitary transformation of the generalized
density can change their value. This is clearly a problem, since these
quantities are directly related to experimental observables.

One must, therefore, choose an adequate representation of the generalized
density in order to compute fragment properties. Obviously, this choice can not
be arbitrary but should instead be guided by physical considerations. The only
inambiguous available experimental data is that fission fragments are
independent of one another: there is no interaction between the two other than
the repulsive Coulomb force. Therefore, the optimal representation should be
the one where $E_{\text{nuc}} \rightarrow 0$. In the HFB approach, this is
achieved if all q.p. are fully localized in a given fragment. Therefore,
for any of the scission configurations chosen in the $I_{q}$ interval 
introduced earlier, physics arguments dictate that one introduces a unitary 
transformation $\mathcal{T}$ of the q.p. such that the localization of each 
individual q.p. is maximized. This transformation would localize the fragments 
by reducing the tails of the densities while leaving the global properties of 
the nucleus unchanged, and would thus ensure that the asymptotic conditions 
of the fission process (the fact that the fission fragments are independent 
systems) are obeyed.

In fact, this need for localization is reminiscent of electronic structure
theory. Similar ideas were introduced long ago in quantum chemistry to describe
the static bonding structure of molecules, see, e.g. Ref.~\cite{(Lev00)} and
references therein. Since the concept of localization is built on the fact that
the wave-function of the system is a product state (of independent particles),
it is also highly relevant to calculations featuring electronic DFT
\cite{(Par95)}; see, e.g. recent developments in ab initio molecular dynamics
\cite{(Sha13),(Ift05),(Par95)}. In all these cases, the original, often called
canonical, calculations tend to yield solutions which are delocalized over the
entire molecule: while such functions can reproduce ionization potentials and
spectral transitions, they fail to describe chemical bonding structure, which
is by nature localized near the atoms \cite{(Edm63)}. By contrast, localized
wave-functions preserve global observables and can also explain chemical
bonding properties.

We choose our unitary transformation $\mathcal{T}$ as follows: for any given 
pair $(\mu,\nu)$ of q.p., we pose
\begin{equation}
\left( \begin{array}{c} U'_{\mu} \\ U'_{\nu} \end{array}\right)
= T
\left( \begin{array}{c} U_{\mu} \\ U_{\nu} \end{array}\right),
\ \ \ 
\left( \begin{array}{c} V'_{\mu} \\ V'_{\nu} \end{array}\right)
= T
\left( \begin{array}{c} V_{\mu} \\ V_{\nu} \end{array}\right),
\end{equation}
with the matrix $T$ of the transformation $\mathcal{T}$ given by
\begin{equation}
T =
\left( \begin{array}{rc}
\cos\theta_{\mu\nu} & \sin\theta_{\mu\nu} \\
-\sin\theta_{\mu\nu} & \cos\theta_{\mu\nu}
\end{array}\right).
\label{eq:unitary}
\end{equation}
The angle of the rotation can be different for every pair of q.p. It can be
chosen so as to maximize the localization of each q.p. of the pair. In the 
following, we drop the indexes $\mu$ and $\nu$ for simplicity, 
$\theta \equiv\theta_{\mu\nu}$. Additional details and discussion can be 
found in Ref.~\cite{(You11),(You14)}.

It is immediate to see that the full density matrix of the compound nucleus 
$\rho$ is invariant under such a transformation. The occupation of any q.p. 
$\mu$, however, becomes
\begin{multline}
N'_{\mu}
=
\sum_{ij}
\left[ 
\cos^{2}\theta V^{*}_{i\mu}V_{j\mu} 
+ 
\sin^{2}\theta V^{*}_{i\nu}V_{j\nu} \right.\\
\left. + \sin\theta\cos\theta ( V^{*}_{i\mu} V_{j\nu} + V^{*}_{i\nu}V_{j\mu} ) \right].
\end{multline}
We write
\begin{equation}
N'_{\mu} = \cos^{2}\theta N_{\mu} + \sin^{2}\theta N_{\nu}
+ \sin\theta\cos\theta\; \omega_{\mu\nu}(-\infty),
\end{equation}
with
\begin{equation}
\omega_{\mu\nu}(z) = \sum_{ij} ( V^{*}_{i\mu} V_{j\nu}
+ V^{*}_{i\nu}V_{j\mu} )d_{ij}(z).
\end{equation}
We note that $\omega_{\mu\nu}(z) = \omega_{\nu\mu}(z)$, and
$\omega_{\mu\nu}(-\infty) = \sum_{i} (V^{*}_{i\mu} V_{i\nu} +
V^{*}_{i\nu}V_{i\mu}) $. For the q.p. $\nu$, the minus sign in front of the
sine in the rotation matrix leads to
\begin{equation}
N'_{\nu} = \cos^{2}\theta N_{\nu} + \sin^{2}\theta N_{\mu}
- \sin\theta\cos\theta\; \omega_{\nu\mu}(-\infty).
\end{equation}

By extension, we find that the occupations of q.p. $\mu$ in each
of the fragment then reads
\begin{equation}
\begin{array}{l}
\displaystyle N'_{1,\mu} = \cos^{2}\theta N_{1,\mu} + \sin^{2}\theta N_{1,\nu} \\
\hfill + \sin\theta\cos\theta\; [\omega_{\mu\nu}(-\infty) - \omega_{\mu\nu}(z_{N})],
\medskip\\
\displaystyle N'_{2,\mu} = \cos^{2}\theta N_{2,\mu} + \sin^{2}\theta N_{2,\nu}
+ \sin\theta\cos\theta\; \omega_{\mu\nu}(z_{N}),
\end{array}
\end{equation}
while for q.p. $\nu$ they are
\begin{equation}
\begin{array}{l}
\displaystyle N'_{1,\nu} = \cos^{2}\theta N_{1,\nu} + \sin^{2}\theta N_{1,\mu} \\
\hfill - \sin\theta\cos\theta\; [\omega_{\nu\mu}(-\infty)-\omega_{\nu\mu}(z_{N})],
\medskip\\
\displaystyle N'_{2,\nu} = \cos^{2}\theta N_{2,\nu} + \sin^{2}\theta N_{2,\mu}
- \sin\theta\cos\theta\; \omega_{\nu\mu}(z_{N}),
\end{array}
\end{equation}
In practice, one determines the optimal angle $\theta$ for each pair by
maximizing the quantity $\ell_{\mu} + \ell_{\nu}$ for the pair $(\mu,\nu)$.


\subsection{Fragment Interaction Energy and Kinetic Energy}
\label{subsec-fragments}

We apply the topological method described in Sec.~\ref{subsec-viz} and the
quantum localization technique presented in Sec.~\ref{subsec-localization} to
the case of $^{240}$Pu. The JCN analysis has identified an interval $I_{q}$ for
the scission configuration,with $q_{N} \leq 0.2$ the most likely candidates
for the actual scission point. For each value of $q_{N}\in I_{q}$, we then
search for the representation of the generalized density yielding the maximum
localization of the fragments by considering all rotations of pairs of q.p.
according to Eq.(\ref{eq:unitary}). In practice, not all pairs of q.p. need to
be rotated: q.p. corresponding to deeply-bound states are pretty well
localized; q.p. with a small occupation contribute little to the interaction
energy, even if they are very delocalized. We can thus limit the computational
burden by applying the localization only on a subset of q.p.. We chose the
following empirical criteria: (i) the occupation of each q.p. $\mu$ is at least
$N_{\mu} > 0.005$, (ii) the localization indicator $\ell_{\mu}$ is
$\ell < 0.75$, and (iii) the q.p. energies of the pair are taken in a 2.0 MeV
energy window, $|E_{\mu} - E_{\nu}| \leq 2$ MeV. In addition, we impose both
q.p. of a given pair to be of the same nature, i.e. either particle-type or
hole-type. We discuss in the appendix how results change with respect to the 
choice of these parameters.

\begin{figure}[!ht]
\centering
\includegraphics[width=\linewidth]{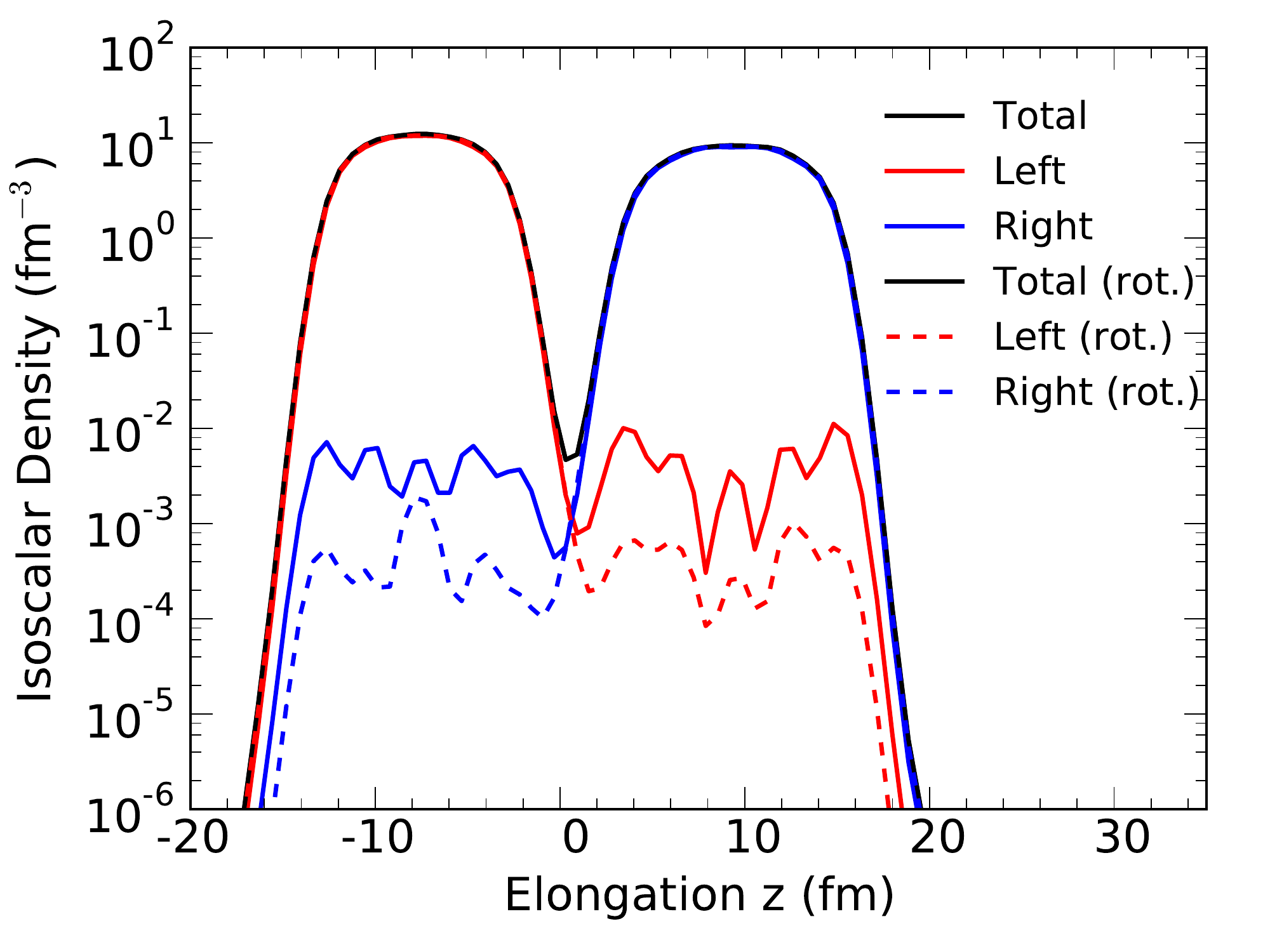}
\caption{(color online) Total, left and right fragment densities before (plain
lines) and after (dashed lines) the localization of q.p. at the $q_{N}=0.4$
point of $^{240}$Pu. Calculations for the SkM* functional.}
\label{fig:density_loc}
\end{figure}

Figure \ref{fig:density_loc} shows the effect of the localization on the total
isoscalar density for the SkM* functional at the point $q_{N}=0.4$.
Superimposed to the total density are the fragment isoscalar densities given by
Eq.~(\ref{eq:density_fragment}). The plain lines correspond to the densities
before localization, the dashed lines after the localization procedure has been
applied. In this example, the localization decreases by about an order of
magnitude the tails of the densities, which will have a sizable impact on the
interaction energy. Note that, as expected, the {\it total} density is
invariant under the unitary transformation (\ref{eq:unitary}).

\begin{figure}[!ht]
\centering
\includegraphics[width=\linewidth]{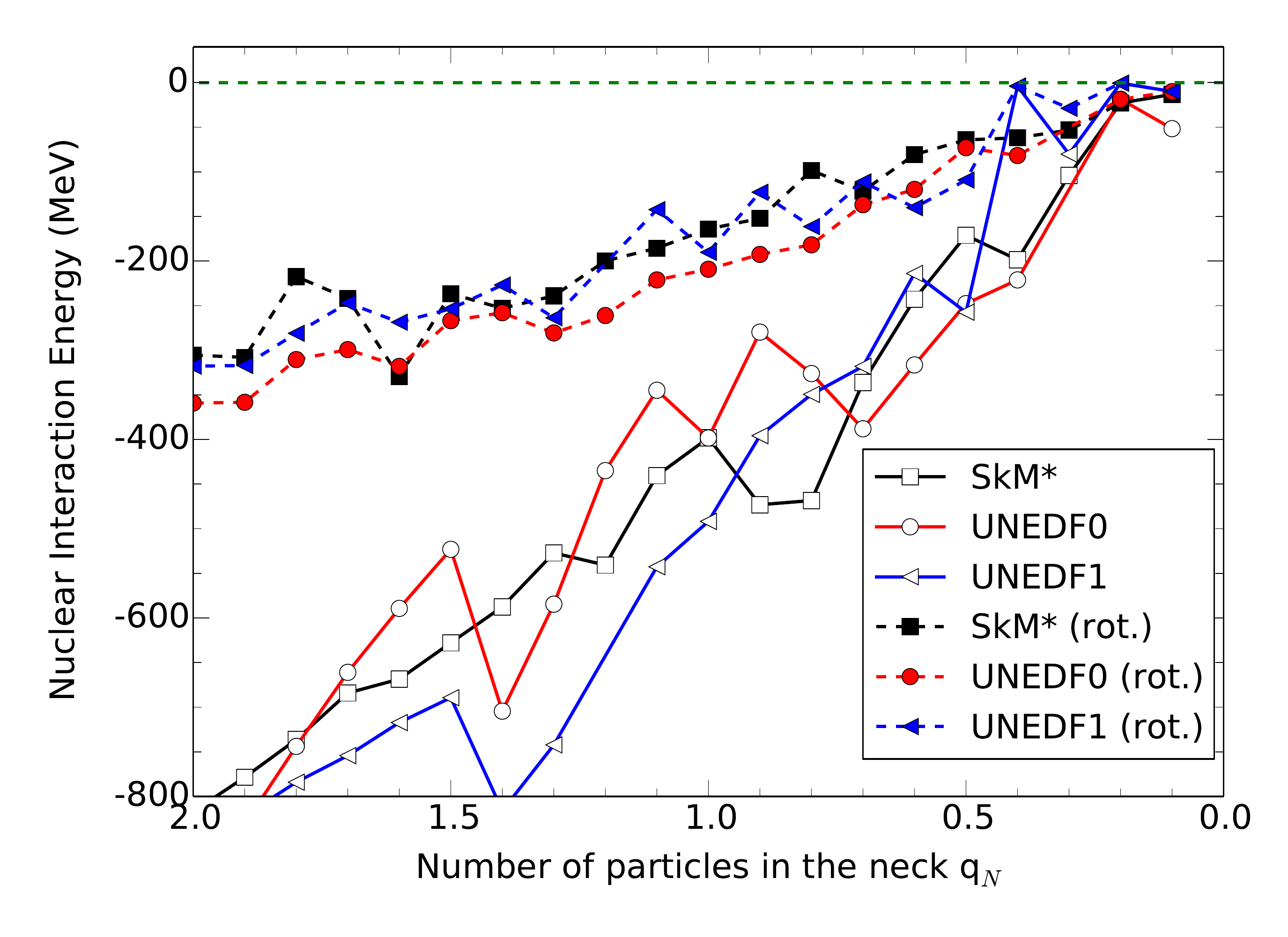}
\caption{(color online) Skyrme interaction energy between the fission fragments
in $^{240}$Pu as a function of the number of particles in the neck for the SkM*,
UNEDF0 and UNEDF1 functionals. Plain curves correspond to the calculation
before the localization is applied, dashed curves to after it has been applied.}
\label{fig:interaction_energy}
\end{figure}

Figure \ref{fig:interaction_energy} shows the nuclear interaction energy betwen
the fragments for the three functionals considered in this work as a function
of the number of particles in the neck. The nuclear interaction energy was 
computed from Eq.~(\ref{eq:interaction_energy}), with the sets (1) and (2) of 
q.p. determined before/after localization. We notice the dramatic effect of the 
localization, especially for larger values of $q_{N}$; it is also worth 
mentioning that the localization tends to average out the fluctuations of 
interaction energy across the range of collective variables. While there are 
small differences between the Skyrme parametrizations, we observe that both 
the overall scale and the trend of the interaction energy as a function of 
$q_{N}$ are similar. The Skyrme interaction energy is also very similar to 
results obtained with the Gogny force \cite{(You11)}. 

\begin{figure}[!ht]
\centering
\includegraphics[width=\linewidth]{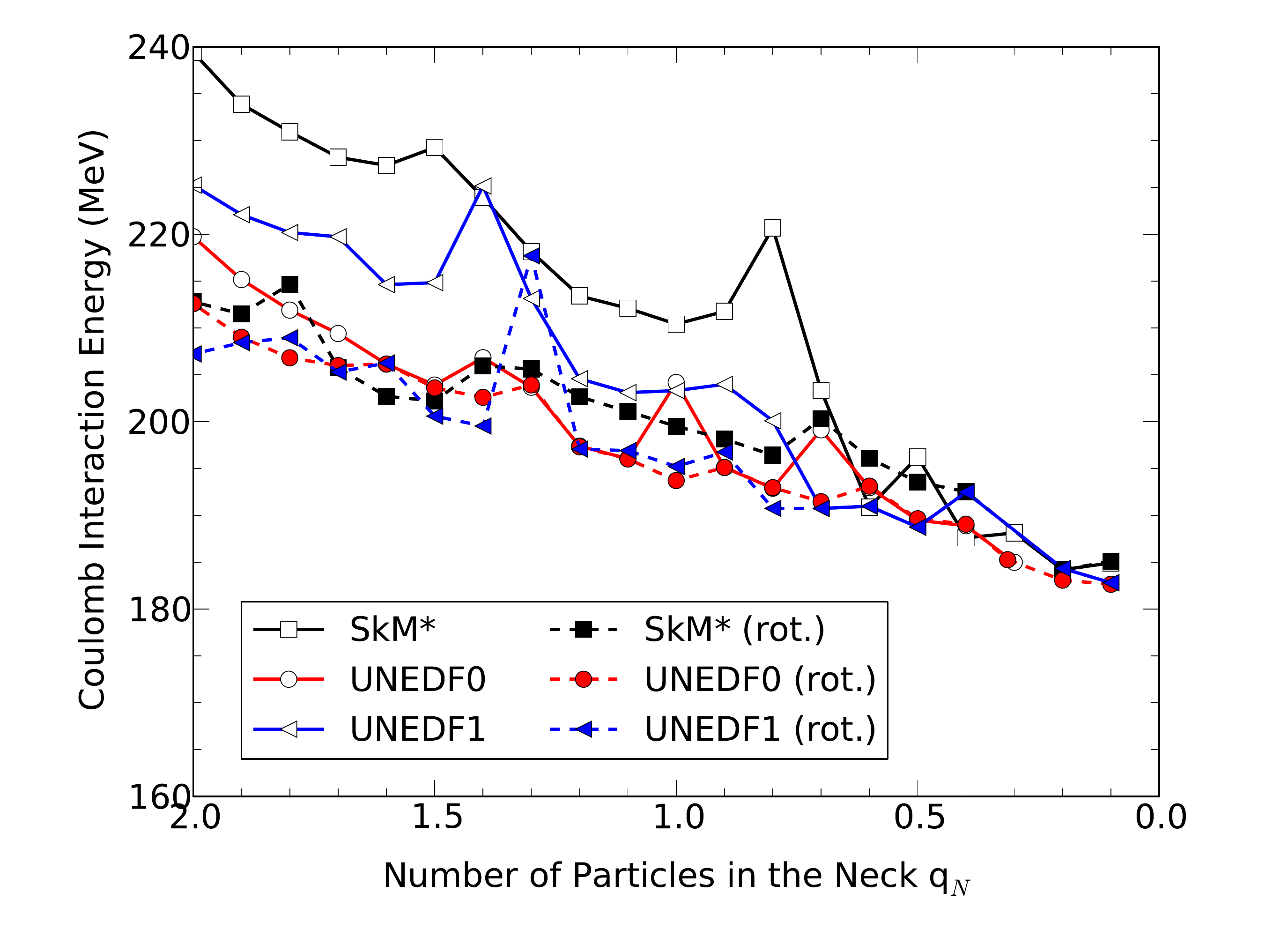}
\caption{(color online) Direct Coulomb interaction energy between the fission
fragments in $^{240}$Pu as a function of the number of particles in the neck
for the SkM*, UNEDF0 and UNEDF1 functionals. Plain curves correspond to the
calculation before the localization is applied, dashed curves to after it has
been applied.}
\label{fig:TKE}
\end{figure}

In the DFT framework, the total kinetic energy (TKE) of the fully accelerated 
fragments is the sum of the Coulomb energy (direct and exchange), the nuclear 
interaction energy, the fragment prescission energy, and the dissipation 
energy,
\begin{equation}
TKE = E_{\text{Cou}}^{(\text{dir})} + E_{\text{Cou}}^{(\text{exc})} 
+ E_{\text{nuc}} +  E_{\text{pre}} + E_{\text{dis}}
\label{eq:TKE}
\end{equation}
In Eq.(\ref{eq:TKE}), the leading term is the the direct part of the Coulomb 
energy. Fig.~\ref{fig:TKE} shows how $E_{\text{Cou}}^{(\text{dir})}$ changes 
as a function of $q_{N}$. Note that we computed this quantity according to 
Eq.(\ref{eq:coulomb_energy_dir}), i.e., fully taking into account the 
deformation of the fragments. At the most likely scission point, $q_{N} = 0.2$, 
see Sec.~\ref{subsec-viz}, the direct Coulomb term is approximately 
$E_{\text{Cou}}^{(\text{dir})} = 185$ MeV and is relatively independent of the 
functional. The exchange contribution $E_{\text{Cou}}^{(\text{exc})}$ is very 
small: it ranges from -4 MeV for large $q_{N}$ values to less than 200 keV 
around $q_{N}=0.2$: it can be neglected at scission. The nuclear interaction 
part depends to a large extent on how well the fragments can be localized. 
In our calculations at $q_{N}=0.2$, we find $E_{\text{nuc}} \approx -22.9$ 
MeV for SkM*, $E_{\text{nuc}} \approx -18.9$ MeV for UNEDF0, and 
$E_{\text{nuc}} \approx -0.7$ MeV for UNEDF1. However, we also observe large 
fluctuations as a function of $q_{N}$ and the parameters of the localization 
procedure, as seen in Figs. \ref{fig:interaction_energy}-\ref{fig:qp_rotation}. 
Conservatively, one may estimate that $E_{\text{nuc}}$ ranges between -25 and 
0 MeV on average, depending on the functional (both particle-hole and 
particle-particle channels), and the quality of the localization. Finally, 
the pre-scission energy is also strongly dependent on the deformation and 
pairing properties of the functional, as discussed in 
Sec.\ref{subsec-skyrmeDependence}; for our small subset of three EDF, it 
ranges between 20 and 30 MeV on average.

Because of internal dissipation, not all the prescission energy is available
to the fragments in the form of kinetic energy. There is a loss equivalent 
to the amount of dissipated energy $E_{\text{dis}}$. Estimating this quantity 
requires to consider the various forms of dissipation. Collective dissipation 
represents the loss of energy due to collective excitations ``transverse'' 
to the fission path: it was estimated to be of the order of 2.1 MeV for 
$Q_{40}$ \cite{(Ber84)} and about 3.1 MeV for $Q_{30}$ \cite{(You12b)}. 
Additional dissipation in the $Q_{22}$ collective variable could also be 
possible, based on the remarks of Sec.~\ref{subsubsec-triaxiality}. 
Intrinsic dissipation could be represented by multi-qp excitations and may 
be estimated in extensions of the GCM framework \cite{(Ber11)}. It does not 
seem unreasonable to estimate that between 5 and 15 MeV of energy could be 
dissipated when combining both collective and intrinsic sources of 
dissipation.

Based on this estimated budget of the various terms in Eq.(\ref{eq:TKE}), 
we can estimate the TKE for the most probable fission in $^{240}$Pu to be 
TKE$\approx 185_{-15}^{+25}$ MeV. This is comparable with what Younes and 
Gogny found using the Gogny force \cite{(You11)}. Although in decent 
agreement with the experimental TKE, which is of the order of 185 MeV for 
the most likely fission \cite{(Tsu00),(Nis95),(Wag84)}, we should point 
out that it is a very conservative and rough estimate based on a sample 
of only three Skyrme functionals (including one, UNEDF0, with notoriously 
poor deformation properties \cite{(Kor10)}). Better constraining 
deformation and pairing properties of functionals should reduce the 
theoretical uncertainty on the amount of pre-scission energy; implementation 
of the localization method on a larger scale, i.e., by considering more 
pairs of q.p., should reduce the fluctuations of the remaining nuclear 
interaction energy. Obtaining reliable estimates of dissipation energy is 
certainly an open question at this time.

We conclude this section by mentioning that at the point of discontinuity in 
Fig.~\ref{fig:PES_1D_energy}, the value of $\hat{Q}_{N}$ is $q_{N} = 4.55$, 
and the value of the Coulomb repulsion energy is 274 MeV: this clearly 
unphysical values justifies {\em a posteriori} the need to include the 
$\hat{Q}_{N}$ degree of freedom (or any collective variable that can 
fulfill its role).


\section{Conclusions}
\label{sec-conclusions}

The description of induced nuclear fission in a microscopic framework based
exclusively on realistic nuclear forces and advanced many-body methods remains
a formidable endeavor. In this work, we have reported some progress in
understanding several of the key ingredients in a theory of fission based on
the nuclear density functional theory with Skyrme energy densities. We have
focused on the benchmark case of the neutron-induced fission of $^{239}$Pu:
\begin{itemize}
\item We have provided a nearly complete mapping of the deformation energy of
the compound nucleus $^{240}$Pu in a 5-dimensional collective space including
all quadrupole degrees of freedom, mass asymmetry, hexadecapole moments and
neck size. While these degrees of freedom are most likely sufficient to cover
the physics of spontaneous fission, where a detailed knowledge of scission is
not really necessary, we point out that the potential energy surface becomes
increasingly complex near scission: further studies of induced fission will
most likely require choosing better sets of collective variables, for example
quantities related to each fragment \cite{(You12a),(Bon07a)}.
\item We have studied the role of triaxiality in the fission process. In
addition to the well-known effect of lowering the first fission barrier heights
of actinides, we have shown that this collective variable also plays a role at
scission. We posit that this extra degree of freedom could contribute to the
decrease of the prescission energy by dissipation in transverse collective
modes, and could result in a broadening of the fission fragment mass yields.
\item We have emphasized the importance of both the form and the
parametrization of the energy functional. Different parameterizations of the
same Skyrme functional can lead to huge fluctuations in deformation energies
\cite{(Nik11)}, which are further magnified near scission; pairing correlations
also play a crucial role in determining the region of scission. 
\item We have presented a general strategy to identify fission fragments in as
automatic a way as possible. This two-step approach first takes advantage of
the Joint Contour Net topological method to define the scission configurations, and
then localizes the fragment following the general idea of Ref.~\cite{(You11)}.
We believe this approach reduces the uncertainty in determining the point where
fission fragment properties must be compared with experimental data. The
application of this technique to the calculation of fission fragment TKE in
$^{239}$Pu(n,f) for the most probable fission shows a decent reproduction of
data.
\end{itemize}

Both the methodology and the results reported in this work pertain to the
static aspects of low-energy fission only. As the excitation energy of the
compound nucleus increases, one should certainly question the capability of
current functionals to capture the physics of fission at the HFB level.
Potential energy surfaces in general, and fission barriers in particular, may
be quite different. One may also wonder if the two-step process introduced here
to define a scission point remains applicable. In the following paper, we will
address these aspects by using a finite-temperature formalism.


\begin{acknowledgments}
We express our deepest gratitude to W. Younes and D. Gogny for many stimulating
and enlightening discussions, and for explaining us the details of their quantum
localization method. We are also thankful to N. Dubray, W. Nazarewicz and J.Pei
for useful comments. Special thanks are addressed to J. McDonnell for a careful
reading of the manuscript. This work was partly performed under the auspices of
the U.S.\ Department of Energy by Lawrence Livermore National Laboratory under
Contract DE-AC52-07NA27344. Funding was also provided by the U.S.\ Department
of Energy Office of Science, Nuclear Physics Program pursuant to Contract
DE-AC52-07NA27344 Clause B-9999, Clause H-9999, and the American Recovery and
Reinvestment Act, Pub. L. 111-5. Computational resources were provided through
an INCITE award ``Computational Nuclear Structure'' by the National Center for
Computational Sciences (NCCS) and National Institute for Computational Sciences
(NICS) at Oak Ridge National Laboratory, and through an award by the Livermore
Computing Resource Center at Lawrence Livermore National Laboratory. Thanks are
also due to the UK Engineering and Physical Sciences Research Council, under
Grant EP/J013072/1.
\end{acknowledgments}


\appendix


\appendix

\section{Characteristics of the localization method}

A unitary transformation $\mathcal{T}$ of the q.p. that do not mix annihilation 
and creation operators transforms the Bobgoliubov matrix $\mathcal{W}$ into 
$\mathcal{W}'$ as
\begin{equation}
\mathcal{W}' = \mathcal{W}
\left( \begin{array}{cc}
T^{T} & 0 \\
0 & T^{\dagger}
\end{array}\right),
\end{equation}
with $T$ a unitary matrix, $TT^{\dagger} = T^{\dagger}T = 1$. Such a unitary 
transformation preserves the HFB equations even though the HFB matrix is not 
diagonal any longer \cite{(Man75)}. This is a simple consequence of the 
Bloch-Messiah theorem \cite{(Blo62)}. Indeed, the HFB matrix in the transformed 
q.p. basis reads
\begin{equation}
\mathcal{H}'
=
\left( \begin{array}{cc}
T^{*} & 0 \\
0 & T
\end{array}\right)
\left( \begin{array}{cc}
E & 0 \\
0 & -E
\end{array}\right)
\left( \begin{array}{cc}
T^{T} & 0 \\
0 & T^{\dagger}
\end{array}\right)
\end{equation}
that is, 
\begin{equation}
\mathcal{H}'
=
\left( \begin{array}{cc}
T^{*}ET^{T} & 0 \\
0 & -TET^{\dagger}
\end{array}\right).
\end{equation}
However, it is straightforward to show that the generalized density 
$\mathcal{R}$ remains invariant under the transformation $\mathcal{T}$, and 
that the HFB equations are preserved,
\begin{equation}
[ \mathcal{H}', \mathcal{R}' ] = 0
\end{equation}
In the particular case where the unitary transformation is defined by 
Eq.(\ref{eq:unitary}), a simple calculation shows that the q.p. energies 
of the pair $(\mu, \nu) $ transform as
\begin{equation}
\left( \begin{array}{cc}
E_{\mu} & 0 \\
0 & E_{\nu}
\end{array}\right) \\
\rightarrow
\left( \begin{array}{cc}
E_{\mu} & 0 \\
0 & E_{\nu}
\end{array}\right)
-
\Delta E \sin\theta
\left( \begin{array}{cc}
\sin\theta & \cos\theta \\
\cos\theta & -\sin\theta
\end{array}\right),
\end{equation}
with $\Delta E = E_{\mu} - E_{\nu}$.


\section{Numerical implementation of the localization method}

\begin{figure*}[!ht]
\centering
\includegraphics[width=\linewidth]{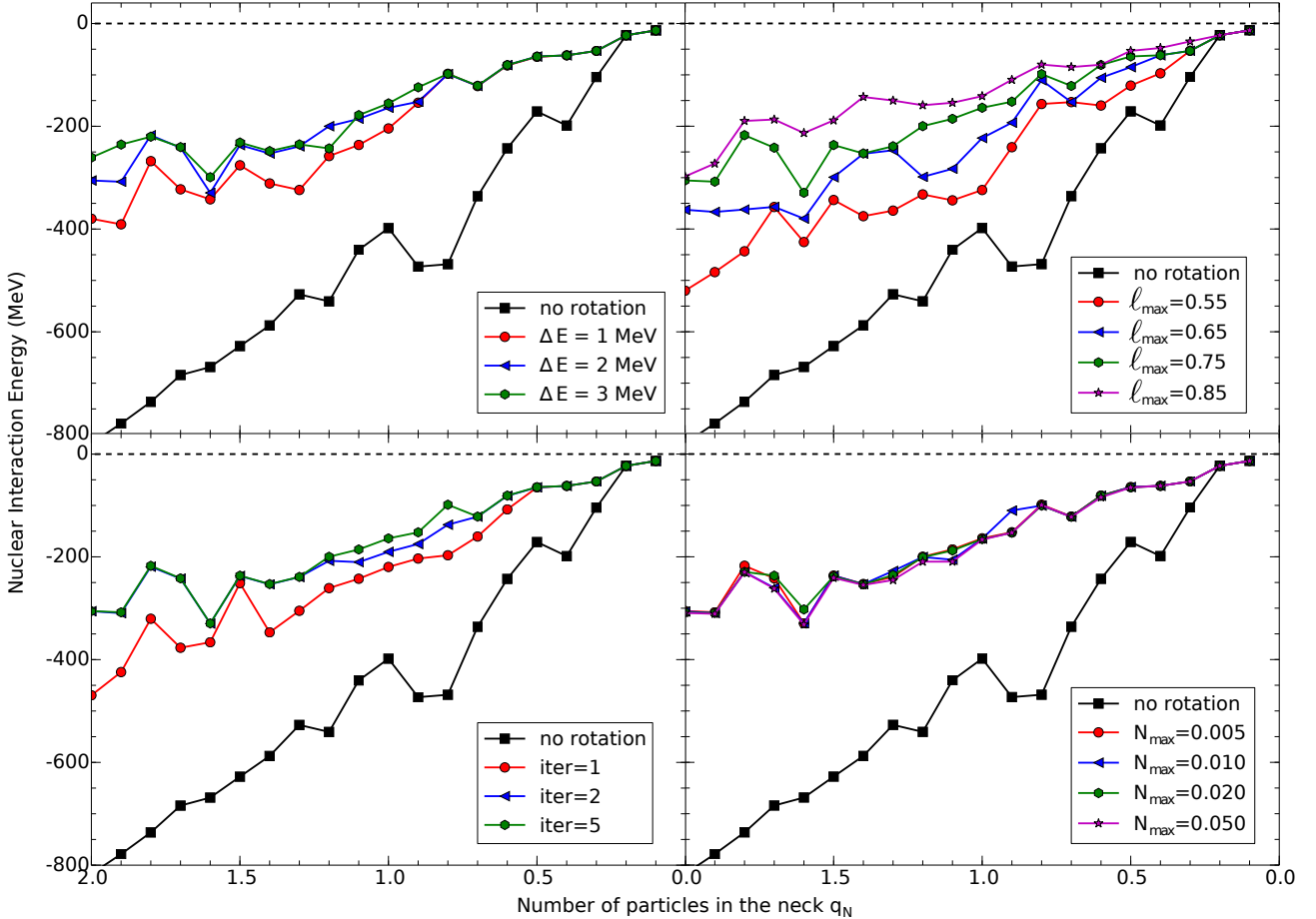}
\caption{(color online) Skyrme interaction energy between fission fragments in 
$^{240}$Pu as a function of the number of particles in the neck for the SkM* 
functional. The parameters of reference for the localization procedure were 
chosen as $\Delta E = 2$ MeV, $\ell_{\text{max}}=0.75$, $N_{\text{max}}=0.005$ 
and 5 iterations. Top left: dependence on $\Delta E$, all other parameters 
being fixed to their reference value; top right: dependence on 
$\ell_{\text{max}}$; bottom left: dependence on the number of iterations; 
bottom right: dependence on $N_{\text{max}}$; See text for further details.}
\label{fig:qp_rotation}
\end{figure*}

As mentioned in Sec.~\ref{subsec-fragments}, the practical implementation of 
the localization procedure depends on a number of parameters. In principle, one 
could consider {\it all} possible pairs of q.p. and rotate the particular 
arrangement of all those pairs that minimizes the interaction energy. The 
computational cost, however, would be formidable. We thus limit the candidates 
to localization by setting various criteria. Figure \ref{fig:qp_rotation} shows 
the dependence of the nuclear interaction energy as a function of these 
parameters. 

In the reference setting that we have adopted, we explore all possible pairs 
of $(\mu, \nu)$ such that $|\Delta E| = |E_{\mu} - E_{\nu}| \leq 2$ MeV, the 
localization of both q.p. is $\ell_{\mu}, \ell_{\nu} \leq 0.75$ and their 
occupation is $N_{\mu}, N_{\nu} \geq 0.005$. In addition, we perform 5 
successive iterations of the localization. Note that, after the first 
iteration, the HFB matrix is not diagonal anymore, hence our first criterion 
cannot based on the eigenvalues $E_{\mu}$, $E_{\nu}$ anymore, but on the 
diagonal elements $E_{\mu\mu}$ and $E_{\nu\nu}$ of the rotated HFB matrix. 

\bibliographystyle{apsrev4-1}
\bibliography{temperature}

\end{document}